\documentclass[referee]{raa}           

\usepackage{graphicx,times}
\usepackage{natbib}
\usepackage{amssymb,amsmath}
\bibpunct{(}{)}{;}{a}{}{,}

\usepackage[pagebackref=true]{hyperref}

\begin{document}

\defcitealias{anders2022}{AKQ22}
\defcitealias{ccm89}{CCM89}
\defcitealias{sfd98}{SFD98}
\defcitealias{schlaflyfinkbeiner2011}{SF11}
\defcitealias{gnilc}{GNILC}
\defcitealias{dbb23}{DBB23}
\defcitealias{chiang2023}{CSFD}
\defcitealias{green2019}{GSZ19}
\defcitealias{gontcharov2023}{GMK23}
\defcitealias{zucker2020}{ZSS20}
\defcitealias{chen2020}{CLY20}

  \title{Foreground Extinction to Extended Celestial Objects -- I. New Extinction Maps}

   \volnopage{Vol.0 (20xx) No.0, 000--000}      
   \setcounter{page}{1}          

\author{G. A. Gontcharov
      \inst{1}
\and A. A. Marchuk
      \inst{1,2}
\and S. S. Savchenko
      \inst{1,2}
\and A. V. Mosenkov
      \inst{3}
\and V.B. Il'in
      \inst{1,2}
\and D. M. Poliakov
      \inst{1}
\and A. A. Smirnov
      \inst{1}
\and H. Krayani
      \inst{2}
   }

\institute{Central (Pulkovo) Astronomical Observatory, Russian Academy of Sciences, Pulkovskoye Chaussee 65/1, 196140 St. Petersburg, Russia; {\it georgegontcharov@yahoo.com}\\
        \and
Saint Petersburg State University, 7/9 Universitetskaya nab., St. Petersburg, 199034 Russia\\
        \and
Astrophysical Research Consortium, c/o Department of Astronomy, University of Washington, Box 351580, Seattle, WA 98195, USA\\
\vs\no
   {\small Received 20xx month day; accepted 20xx month day}}

\abstract{We present a new two-dimensional (2D) map of total Galactic extinction, $A_\mathrm{V}$, across the entire dust half-layer from the Sun to extragalactic space for Galactic latitudes $|b| > 13\degr$, as well as a three-dimensional (3D) map of $A_\mathrm{V}$ within 2~kpc of the Sun.
These maps are based on $A_\mathrm{V}$ and distance estimates derived from a dataset, which utilizes {\it Gaia} Data Release 3 parallaxes and multi-band photometry for nearly 100 million dwarf stars. We apply our own corrections to account for significant systematics in this dataset. Our 2D map achieves an angular resolution of 6.1~arcmin, while the 3D map offers a transverse resolution of 3.56~pc --- corresponding to variable angular resolution depending on distance --- and a radial resolution of 50~pc. In constructing these maps, we pay particular attention to the solar neighborhood (within 200~pc) and to high Galactic latitudes.
The 3D map predicts $A_\mathrm{V}$ from the Sun to any extended object within the Galactic dust layer with an accuracy of $\sigma(A_\mathrm{V}) = 0.1$~mag. The 2D map provides $A_\mathrm{V}$ estimates for the entire dust half-layer up to extragalactic distances with an accuracy of $\sigma(A_\mathrm{V}) = 0.07$~mag. We provide $A_\mathrm{V}$ estimates from our maps for various classes of extended celestial objects with angular size primarily in the range of 2--40~arcmin, including 19,809 galaxies and quasars, 170 Galactic globular clusters, 458 open clusters, and several hundreds molecular clouds from two lists. We also present extinction values for 8,293 Type Ia supernovae. Comparison of our extinction estimates with those from previous maps and literature sources reveals systematic differences, indicating large-scale spatial variations in the extinction law and suggesting that earlier 2D reddening maps based on infrared dust emission tend to underestimate low extinction values.
}

   \authorrunning{G. Gontcharov et al. }            
   \titlerunning{Foreground Extinction to Celestial Objects}  
   \maketitle

%
\section{Introduction}           
\label{sect:intro}

The spatial distribution of dust, as inferred from stellar reddening and interstellar extinction, plays a crucial role in studying the properties of celestial objects, as well as the structure and evolution of both our Galaxy and extragalactic systems. The individual reddening or cumulative extinction toward a celestial object can be most accurately determined from its spectral energy distribution (SED). However, SEDs have been observed and analyzed for only a small fraction of stars, even within the nearest regions of the Milky Way. 
Nevertheless, individual reddening or extinction estimates can be used to construct maps that smooth out the natural small-scale fluctuations in the dust medium from star to star and capture large-scale spatial variations in reddening or extinction in tabulated form. Alternatively, these large-scale variations can be described using analytical models. 
Both reddening/extinction maps and models can then be used to estimate extinction for any celestial object with known coordinates.

A three-dimensional (3D) map represents reddening or extinction as a function of Galactic longitude $l$, latitude $b$, and distance $R$ from the Sun, or equivalently, in terms of the rectangular Galactic coordinates $XYZ$.\footnote{We adopt a Galactic rectangular coordinate system centered on the Sun, with the $X$, $Y$, and $Z$ axes pointing toward the Galactic center, in the direction of Galactic rotation, and toward the North Galactic Pole (NGP), respectively (similarly, SGP denotes the South Galactic Pole). These coordinates are calculated from $R$, $l$, and $b$.}

For distant celestial objects --- such as galaxies, quasars, Type Ia supernovae (SN~Ia), and globular clusters in the Galactic halo --- a two-dimensional (2D) map is sufficient to provide the total Galactic extinction (TGE) and reddening across the entire dust layer from the Sun to these objects, as a function of Galactic coordinates $l$ and $b$ only.
The most widely used 2D map is that of \citet[][hereafter SFD98]{sfd98}, which is based on data from the {\it Cosmic Background Explorer (COBE)} and the {\it Infrared Astronomical Satellite (IRAS)}. This map has been refined by \citet[][hereafter SF11]{schlaflyfinkbeiner2011} and \citet[][hereafter CSFD]{chiang2023}. Another widely used map is that of \citet[][hereafter GNILC]{gnilc}, constructed from observations by the {\it Planck Space Observatory}. These 2D maps are based on measurements of dust infrared emission along the entire line of sight (LOS), followed by a calibration between dust emission and reddening.

The {\it Gaia} mission \citep{gaiadr3a1} has led to significant advances in the study of interstellar dust, extinction, and reddening in the Milky Way, particularly by enabling the construction of three-dimensional (3D) maps using its precise parallax measurements. 
Also, {\it Gaia} parallaxes can be used to refine 2D extinction maps, as they provide an upper limit on the extinction along each LOS, effectively representing the asymptotic value of corresponding 3D maps.

Since the publication of the first 3D extinction map based on {\it Gaia} parallaxes \citep{gontcharov2017a}, considerable efforts have been devoted to producing numerous 3D maps. Among these, the widely used map by \citet[][hereafter GSZ19]{green2019} stands out for its accuracy and broad spatial coverage, extending from approximately 200~pc to several kiloparsecs from the Sun and encompassing three-quarters of the sky.

Numerous 2D and 3D extinction maps have been compared and analyzed in detail, for example, by \citet[][and references therein]{gontcharov2021a,gontcharov2021b}. These studies conclude that the total uncertainty --- comprising both statistical and systematic components --- of any state-of-the-art 2D or 3D extinction map is, at best, $\sigma(A_\mathrm{V}) = 0.08$~mag. This level of uncertainty is comparable to the typical extinction values $A_\mathrm{V}$ in the $V$ band near the Sun and at high Galactic latitudes.\footnote{We highlight the region within approximately 200~pc of the Sun as one where the typical uncertainty in reddening/extinction is comparable to the extinction itself. Moreover, the low stellar density in this region often prevents the application of certain reddening/extinction estimation methods, such as that used by \citetalias{green2019}.}
Furthermore, natural fluctuations in the interstellar dust medium occur on spatial scales larger than at least 0.1~pc \citep[][and references therein]{panopoulou2022}, introducing additional uncertainty into the predictions of any 2D or 3D map when applied to point sources. Since these maps inherently smooth over small-scale variations, they cannot capture such a fine structure. The resulting uncertainty ranges from approximately $\sigma(A_\mathrm{V}) = 0.06$~mag at high Galactic latitudes to $\sigma(A_\mathrm{V}) = 0.33$~mag or higher near the Galactic plane and within dense dust clouds exhibiting steep extinction gradients \citep[][and references therein]{green2015,gontcharov2019,gontcharov2022}.
Considering the typical amplitude of these fluctuations, the inherent uncertainties in extinction maps, and the typical uncertainties in SED-based individual $A_\mathrm{V}$ estimates, one can conclude that map-based predictions are generally preferable to individual estimates only at high latitudes or for extended celestial objects whose angular sizes are comparable to the map’s resolution (see discussion by \citealt[][hereafter GMK23]{gontcharov2023}). For larger extended objects, extinction maps should be used to analyze spatial variations in extinction across their extent.
The typical angular resolution of modern extinction maps ranges from 3 to 20 arcminutes. In this study, we adopt an angular resolution of 6.1 arcminutes for our 2D map. Accordingly, we define extended objects as those with angular diameters between 2 and 40 arcminutes, for which extinction can be reasonably approximated by a single value from our map without requiring analysis of internal extinction variation.

It is evident that new 2D and 3D extinction maps --- more accurate particularly in the solar neighborhood and at high Galactic latitudes, and based on individual extinction estimates combined with {\it Gaia} parallaxes --- are needed, especially for predicting extinction toward extended and high-latitude objects. Precise extinction estimates are crucial for the study of extragalactic systems, as even small uncertainties in foreground extinction can significantly affect measurements of galaxy colors, surface brightness profiles, and SEDs. This, in turn, can bias derived physical properties such as stellar masses, star formation rates, and dust content in galaxies under study. Furthermore, high-latitude fields are often used as reference regions for cosmological surveys and low-surface-brightness galaxy searches, where an accurate correction for Galactic dust is essential to avoid systematic errors.

In this study, we present such 2D and 3D maps, constructed using $A_\mathrm{V}$ and distance ($R$) estimates from \citet[][hereafter AKQ22]{anders2022}\footnote{\url{https://data.aip.de/projects/starhorse2021.html} or \url{https://cdsarc.cds.unistra.fr/viz-bin/cat/I/354}}. As noted by \citetalias{anders2022}, ``In principle, our extinction results can be used to infer precise distances to individual dust clouds and to infer the three-dimensional distribution of dust.'' We adopt this approach in the present work.
Because the \citetalias{anders2022} dataset samples the full extent of the Galactic dust layer at mid and high Galactic latitudes, it enables the construction of both 2D and 3D extinction maps.

Finally, we provide $A_\mathrm{V}$ estimates from our maps for selected samples of SN~Ia and extended celestial objects --- including galaxies and quasars, Galactic globular clusters, open clusters, and molecular clouds --- and compare these estimates with those from widely used extinction maps and values reported in the literature.

The remainder of this paper is organized as follows.
In Sect.~\ref{sect:data}, we present the data used in this study.
Systematic effects in the dataset are analyzed in Sect.~\ref{sect:systematics}.
In Sect.~\ref{sect:maps}, we describe the construction of our extinction maps and highlight the improvements over our previous maps presented in \citetalias{gontcharov2023}.
Sect.~\ref{sect:tests} is devoted to testing our maps and providing extinction estimates for galaxies and quasars, SN~Ia, globular clusters, open clusters, and molecular clouds.
Our main findings and conclusions are summarized in Sect.~\ref{sect:conclusions}.
An additional comparison of various extinction maps is presented in Appendix~\ref{addcmds}.

\section{Data}           
\label{sect:data}

In this study, we use the dataset from \citet[][hereafter AKQ22]{anders2022}, which provides individual extinction estimates, distances, and stellar parameters (including age, mass, effective temperature, metallicity, and surface gravity) for several hundred million stars within a few kiloparsecs of the Sun. To date, this represents one of the most extensive and precise datasets available for studies of the Milky Way. Their analysis combines parallaxes and photometry from {\it Gaia} Data Release 3 (DR3) with multi-band photometry from several large-scale surveys: the Two Micron All-Sky Survey (2MASS; \citealt{2mass}), the {\it Wide-field Infrared Survey Explorer} ({\it WISE}; \citealt{wise}), the Panoramic Survey Telescope and Rapid Response System Data Release 1 (Pan-STARRS, PS1; \citealt{chambers2016}), and the SkyMapper Southern Sky Survey Data Release 2 (SMSS DR2; \citealt{onken2019}). To derive the most probable stellar parameters, \citetalias{anders2022} employ the StarHorse code \citep{queiroz2018}, which fits theoretical PARSEC1.2S+COLIBRIS37 isochrones \citep{bressan2012} to the observed data in color--magnitude diagrams (CMDs). These isochrones are computed for the solar metallicity scale and do not account for $\alpha$-element enhancement, which may limit accuracy in Galactic halo populations. \citetalias{anders2022} adopt prior assumptions on the geometry, metallicity, and age distributions of the main Galactic components. Notably, for the region of the sky covered by PS1, their extinction prior is based on \citetalias{green2019}, while for the remaining quarter of the sky, they rely on the 3D extinction model of \citet{drimmel2003}.
The typical precision of the $A_\mathrm{V}$ estimates is approximately 0.15~mag for bright stars and 0.20~mag for faint stars.

Our first attempt to construct 2D and 3D extinction maps based on the \citetalias{anders2022} dataset was presented in \citetalias{gontcharov2023}. Using the $R$ and $A_\mathrm{V}$ estimates from \citetalias{anders2022} for approximately 100 million dwarf stars within 2.5~kpc of the Sun, we produced a set of extinction-related maps with the following key components:
(1) 3D maps of $A_\mathrm{V}$ and $A_\mathrm{G}$ (extinction in the {\it Gaia} $G$ filter) within 2~kpc;
(2) a 3D differential extinction map (representing the dust spatial density distribution) in the same volume;
(3) a 3D map of variations in the extinction law (i.e., the wavelength dependence of extinction) within 800~pc of the Sun; and
(4) a 2D map of TGE for intermediate and high Galactic latitudes, specifically for $|b| > 13\degr$.
The lower limit of $|b| = 13\degr$ arises from estimates of the vertical extent of the Galactic dust layer. For instance, \citet{gontcharov2021b} showed that extinction growth becomes negligible beyond $|Z| \approx 450$~pc from the Galactic mid-plane. Therefore, even when restricting extinction measurements to within 2~kpc from the Sun, we can reliably approximate the full-column TGE for latitudes $|b| > \arcsin(450/2000) \approx 13\degr$.

\citetalias{gontcharov2023} used dwarf stars --- rather than giants, as employed by \citet[][hereafter DBB23]{dbb23} for their 2D map based on the same \citetalias{anders2022} dataset --- because the dwarf sample is significantly more complete in the solar neighborhood and provides photometry of higher fidelity.
The 3D maps in \citetalias{gontcharov2023} have a radial resolution of 50~pc and a transverse resolution ranging from 3.6 to 11.6~pc. The 2D map has an angular resolution of 6.1 arcminutes, consistent with that of commonly used maps by \citetalias{sfd98}, \citetalias{schlaflyfinkbeiner2011}, and \citetalias{chiang2023}.
The reported uncertainty in the \citetalias{gontcharov2023} maps is $\sigma(A_\mathrm{V}) = 0.06$~mag, although the present study indicates that the true uncertainty may be slightly larger. A major contributor to this uncertainty is a nonphysical systematic trend in the \citetalias{anders2022} data --- specifically, a dependence of $A_\mathrm{V}$ on distance $R$ of up to $\Delta A_\mathrm{V} = \pm0.04$~mag (see figures~1 and 2 in \citetalias{gontcharov2023}). This systematic effect manifests as an artificial increase in $A_\mathrm{V}$ with distance along many LOSs, which hampers the reliable construction of extinction maps based on the \citetalias{anders2022} dataset.
To mitigate this issue, \citetalias{gontcharov2023} applied an empirical correction to $A_\mathrm{V}$, modeled as a sine function of distance modulus. While the origin of this systematic trend was unclear, it was suggested that it might result from an inadequate treatment of stellar metallicity or the presence of unresolved binary systems.
Now the origin of this systematics seems to be established and discussed in Sect.~\ref{sect:systematics}.

The construction of the 2D and 3D extinction maps in this study generally follows the methodology developed in \citetalias{gontcharov2023}, with several updates described in subsequent sections. The input data were prepared and selected by \citetalias{anders2022}, who followed the recommendations of the original data providers for {\it Gaia} DR3, SMSS, PS1, 2MASS, and {\it WISE}. In addition, \citetalias{anders2022} applied a number of quality criteria to ensure the reliability of the derived parameters. Building on this, we applied the following selection criteria to construct our working sample:
\begin{itemize}
    \item \verb"dist50<3.0": stars within 3~kpc of the Sun (beyond this limit, the sample becomes strongly incomplete, introducing significant biases);
    \item \verb"fidelity>0.5": ensuring reliable astrometric solutions;
    \item \verb"sh_outflag='0000'": selecting stars with the highest fidelity in the StarHorse output parameters;
    \item \verb"(av84-av16)/2<0.25": selecting stars with extinction uncertainties better than 0.25~mag;
    \item \verb"(dist84-dist16)/2/dist50<0.25": ensuring relative distance uncertainties better than 25\%.
\end{itemize}

We also followed the recommendation of \citetalias{anders2022} to apply a cut on \verb"bp_rp_excess_corr" to minimize the impact of background flux from nearby sources on {\it Gaia} photometry.
To isolate dwarf stars, we imposed additional constraints: \verb"logg50 > 3.95" (on surface gravity) and \verb"mg0 > 3.3" (on absolute magnitude in the $G$ band).
It is important to emphasize the role of the \verb"sh_outflag" parameter. We found that stars with high-quality StarHorse outputs (\verb"sh_outflag = '0000'") show significantly different $A_\mathrm{V}$-–$R$ trends compared to lower-fidelity stars. Although the latter represent a minor fraction of the sample, they contribute noticeably to the systematic trend of increasing $A_\mathrm{V}$ with distance discussed earlier. This finding aligns with the recommendation in \citetalias{anders2022}: ``unproblematic results from the point of view of StarHorse can thus be filtered by requiring \verb"sh_outflag = '0000'".''

Our final sample consists of 107,114,524 dwarf stars located within 3~kpc of the Sun.

The most significant update relative to \citetalias{gontcharov2023} is the systematic investigation and correction of biases present in the data, as described in the following section.

\section{Systematics}
\label{sect:systematics}

\begin{figure} 
   \centering
   \includegraphics[width=14.0cm, angle=0]{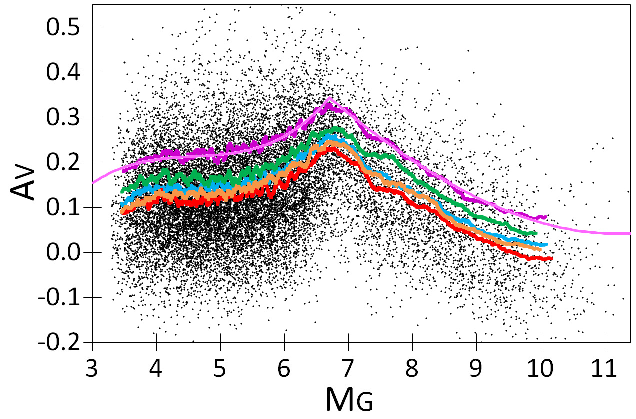}
   \caption{Moving average of $A_\mathrm{V}$ over 301 data points as a function of absolute magnitude $M_G$ for spatial cones centered around five directions, shown by colored curves:
red — North Galactic Pole (NGP),
blue — South Galactic Pole (SGP),
orange — $(l=180\degr,\ b=+45\degr)$,
green — $(l=270\degr,\ b=+50\degr)$,
purple — $(l=90\degr,\ b=-60\degr)$.
For illustration, the original unaveraged data for the NGP direction are shown as black symbols.
The light purple curve represents the systematic trend in the $(l=90\degr,\ b=-60\degr)$ direction corrected using Eqs.~(\ref{correction1}) and (\ref{correction2}).}
   \label{fig:mg0_av.eps}
   \end{figure}

\begin{figure} 
   \centering
   \includegraphics[width=14.0cm, angle=0]{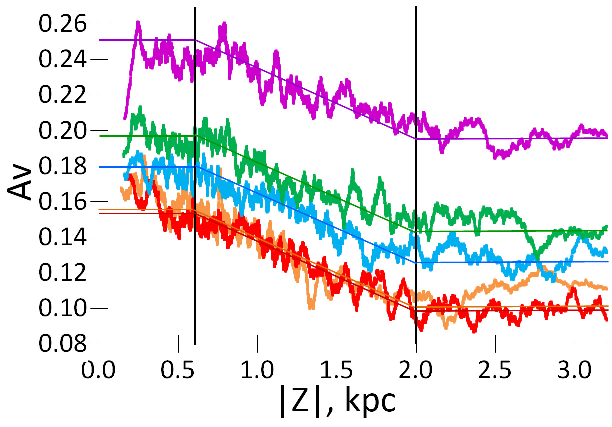}
\caption{Moving average of $A_\mathrm{V}$ over 301 data points as a function of $|Z|$, after accounting for the dependence of $A_\mathrm{V}$ on absolute magnitude $M_G$, for the same five spatial cones shown in Fig.~\ref{fig:mg0_av.eps}, indicated by the same colors.
The colored straight lines represent the systematic trends corrected using Eqs.~(\ref{second1}) and (\ref{second2}).} 
   \label{fig:z_av.eps}
   \end{figure}

In this section, we present a detailed analysis of the systematic effects present in the \citetalias{anders2022} dataset.

Unresolved binaries must be ruled out as the primary cause of the $R$ versus $A_\mathrm{V}$ systematic trend observed in the \citetalias{anders2022} data, as shown by the analysis of non-single stars in \citet{gaiadr3a34}. Specifically, unresolved binaries constitute less than 10\% of nearby stars in {\it Gaia} DR3, and their numbers drop sharply beyond 200~pc --- yet the systematic trend persists well beyond this distance. Therefore, both the number and the relative fraction of unresolved binaries are insufficient to account for the observed systematics in the \citetalias{anders2022} data.
However, we have identified three alternative sources contributing to this effect.

The first reason is well established in recent studies by \citet{heyl2022,brandner2023a,brandner2023b,brandner2023c,wang2025}, which demonstrate that various sets of theoretical isochrones, including PARSEC isochrones used by \citetalias{anders2022}, systematically deviate from observations of open clusters in the dwarf regime of CMDs. A likely explanation for this discrepancy is inaccuracies in the modeling of low-mass stars \citep[][and references therein]{wang2025}. This effect becomes more pronounced for stars with absolute magnitudes $M_G > 10$~mag. Fortunately, our sample includes very few such stars, as they are largely excluded during the data cleaning process. The small number of remaining stars with $M_G > 10$~mag --- found within 650~pc of the Sun --- are accounted for through a dedicated correction, as described later in this section.

This isochrone-to-data mismatch gives rise to a systematic trend between $M_G$ and $A_\mathrm{V}$ in our dataset, as illustrated in Fig.~\ref{fig:mg0_av.eps}. The figure shows moving average curves for various spatial cones, along with individual data points for stars in the cone toward the NGP. This systematic pattern closely resembles that in figure 2 of \citet{brandner2023c}, though it appears inverted relative to our presentation.
Fig.~\ref{fig:mg0_av.eps} further demonstrates that this pattern is consistent across many LOSs, differing only by a constant offset in $A_\mathrm{V}$ --- that is, a vertical shift in the pattern --- specific to each LOS.

Applying an empirical correction for this type of systematic trend is a common approach \citep{wang2025}. In our case, we adopt a correction as a function of absolute magnitude $M_G$, with a break point at $M_G = 6.7$~mag and a fixed average value of $A_\mathrm{V}$. The correction is given by the following polynomial expressions:
\begin{equation}
\Delta A_\mathrm{V}=-0.0119\,M_G^3+0.1634\,M_G^2-0.7557\,M_G+1.1859,\,\mbox{ if  } M_G<6.7,
\label{correction1}
\end{equation}
\begin{equation}
\Delta A_\mathrm{V}=-0.0008\,M_G^3+0.0083\,M_G^2+0.1148\,M_G-1.0223,\,\mbox{ if  } M_G>6.7,
\label{correction2}
\end{equation}
where the coefficients are determined using the least squares fitting method.

The second contributor to the $R$ versus $A_\mathrm{V}$ systematics in the \citetalias{anders2022} data is likely the assignment of an incorrect metallicity to the best-fitting isochrone, as previously suggested by \citetalias{gontcharov2023}. This issue arises from the use of broad- and intermediate-band photometry, which is only weakly sensitive to metallicity, as noted by \citetalias{anders2022}.
Combined with the well-known degeneracy between metallicity and extinction --- i.e., the difficulty in distinguishing whether a star appears redder due to higher metallicity or higher extinction --- this limitation can lead to systematic errors in the estimated extinction values.

Fig.~\ref{fig:z_av.eps} shows the moving average of $A_\mathrm{V}$ as a function of vertical distance from the Galactic mid-plane, $|Z|$, for various spatial cones. This analysis is performed after applying the $M_G$-dependent corrections given by Eqs.~(\ref{correction1}) and (\ref{correction2}).
As with the $M_G$ versus $A_\mathrm{V}$ systematic trend, the $|Z|$ versus $A_\mathrm{V}$ pattern appears consistent across different LOSs, differing only by a constant offset in $A_\mathrm{V}$ --- resulting in a vertical shift of the overall trend.
Given the previously mentioned degeneracy between metallicity and extinction, we attribute this behavior to an inaccurate metallicity gradient with $|Z|$ as adopted in the priors of \citetalias{anders2022}.
The colored straight lines in Fig.~\ref{fig:z_av.eps} represent the empirical correction we apply:
\begin{equation}
\Delta A_\mathrm{V}=-0.08+0.00004\,|Z|,\,\mbox{\;if \;} 600<|Z|<2000\mbox{ pc,}
\label{second1}
\end{equation}
\begin{equation}
\Delta A_\mathrm{V}=-0.056,\,\mbox{\;if \;} |Z|<600\mbox{ pc},
\label{second2}
\end{equation}
where the coefficients are determined by the least squares method.
For $|Z|>2$~kpc, the $A_\mathrm{V}$ estimates from the dwarf sample are consistent with those obtained from a control sample of giants selected for comparison. Therefore, no correction is applied in this region.

\begin{figure} 
   \centering
   \includegraphics[width=14.0cm, angle=0]{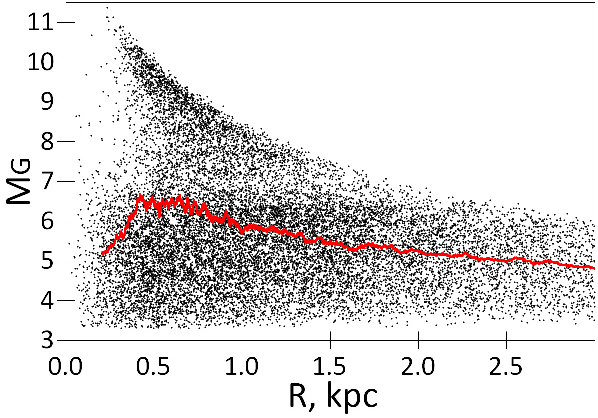}
\caption{Absolute magnitude $M_G$ as a function of distance $R$ for stars in our sample located within a $4\degr$ cone around the SGP.
The red curve shows the moving average computed over 249 points.} 
   \label{fig:z_mg0.eps}
   \end{figure}

\begin{figure} 
   \centering
   \includegraphics[width=15.0cm, angle=0]{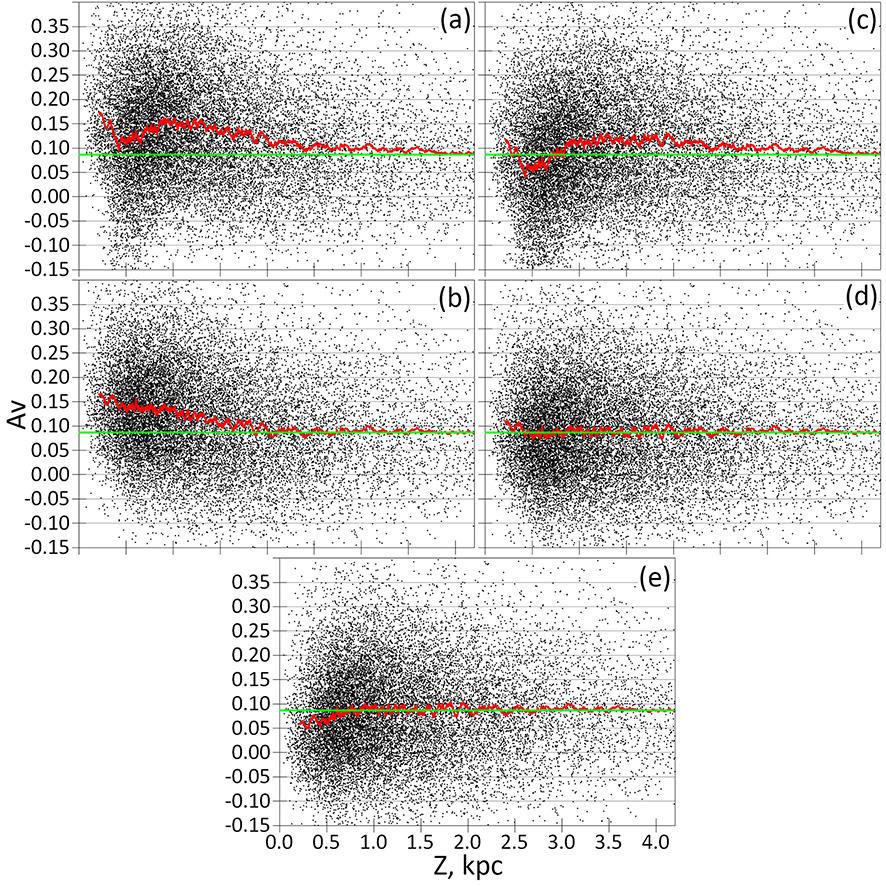}
   \caption{Total Galactic extinction $A_\mathrm{V}$ as a function of vertical distance $Z$ for \citetalias{anders2022} dwarf stars within a $4\degr$ cone around the NGP:
(a) original values;
(b) after applying the $M_G$-dependent correction [Eqs.(\ref{correction1}) or (\ref{correction2})];
(c) after applying the $|Z|$-dependent correction [Eqs.(\ref{second1}) or (\ref{second2})];
(d) after applying both corrections;
(e) after applying the final correction [Eq.~(\ref{third})].
The red curve represents the moving average over 249 points.
The green line indicates the TGE in the direction of the NGP.
The sample is extended to 4.2~kpc for illustrative purposes.}
   \label{fig:corrections.eps}
   \end{figure}

The third contributor to the $R$ versus $A_\mathrm{V}$ systematics in the \citetalias{anders2022} data is evident in Fig.~\ref{fig:z_mg0.eps}, which shows $M_G$ as a function of $R$ for stars within a $4\degr$ cone around the SGP. A similar pattern is observed for all LOSs.
The figure shows that faint stars gradually drop out of the sample with increasing distance due to selection effects. For $R > 650$~pc, this selection leads to a relatively monotonic trend in $M_G$ as a function of $R$, as reflected in the moving average of $M_G$. This behavior is already accounted for in the $R$ versus $A_\mathrm{V}$ systematics via the $M_G$ versus $A_\mathrm{V}$ correction applied through Eqs.~(\ref{correction1}) and (\ref{correction2}).
However, for $R < 650$~pc, this trend changes due to the preferential removal of nearby faint stars. This is primarily a result of our selection criterion \verb"sh_outflag='0000'", which excludes stars with low-fidelity StarHorse output parameters. As seen in Fig.~\ref{fig:z_mg0.eps}, stars in the range $7 < M_G < 11$ --- which typically have lower $A_\mathrm{V}$ values (as shown in Fig.~\ref{fig:mg0_av.eps}) --- are progressively eliminated as $R$ decreases from 650~pc to 0~pc. This leads to a bias in the average $A_\mathrm{V}$ within $R < 650$~pc: the closer the spatial point, the more likely its extinction is overestimated due to the absence of low-$A_\mathrm{V}$ stars.

Given that $A_\mathrm{V}=0$ at $R=0$, we adopt the following linear correction to account for this bias, with coefficients determined via least squares fitting:
\begin{equation}
\Delta A_\mathrm{V}=-0.065+0.0001\,R,\,\mbox{\;if \;} R<650\mbox{ pc,}
\label{third}
\end{equation}
where the coefficients are determined using the least squares fitting method.

Finally, Fig.~\ref{fig:corrections.eps} illustrates the elimination of the $R$ versus $A_\mathrm{V}$ systematics after applying our corrections, shown for the spatial cone toward the NGP. The figure demonstrates that only the combined application of all three corrections successfully suppresses the significant systematic trends and yields a physically meaningful dependence of $A_\mathrm{V}$ on $R$. In particular, $A_\mathrm{V}$ increases with distance within the Galactic dust layer up to approximately $Z \approx 500$~pc, beyond which it remains nearly constant.

\section{Creating Maps}
\label{sect:maps}

To construct our 2D and 3D extinction maps, we compute averages of individual $A_\mathrm{V}$ estimates within defined angular cells (for the 2D map) and spatial bins (for the 3D map), respectively.

Unlike \citetalias{gontcharov2023}, where the 2D map was limited to $|b| > 13\degr$, we calculate our 2D extinction map for all Galactic latitudes. However, the method of estimating $A_\mathrm{V}$ differs between high and low latitudes. For $|b| > 13\degr$, we follow the previous approach, averaging $A_\mathrm{V}$ values for stars with distances $R > R_{\mathrm{limit}}$, where $R_{\mathrm{limit}} \equiv 450/\sin{|b|}$ pc. For $|b| < 13\degr$, we adopt the $A_\mathrm{V}$ value in the farthest distance bin (at 2~kpc) as the 2D map estimate. This choice reflects the challenges at low latitudes, where $A_\mathrm{V}$ estimates at larger distances are strongly affected by fluctuations in the dust distribution and by the loss of stars due to heavy extinction from dense dust clouds.

The quantity $R_{\mathrm{limit}}$ defines the distance beyond which we no longer consider spatial variations in $A_\mathrm{V}$. For $R > R_{\mathrm{limit}}$, we assume that such variations are negligible at high Galactic latitudes ($|b| > 13\degr$), while at low latitudes ($|b| < 13\degr$), we advise using our maps with caution due to increased uncertainty. The value of $R_{\mathrm{limit}}$ varies from 450~pc near the Galactic poles to 2~kpc at the Galactic equator.

We adopt a uniform transverse grid (in Galactic longitude $l$ and latitude $b$) for both our 2D and 3D maps, with a resolution of 6.1 arcminutes. This represents a significant improvement over the 20 arcminute step used in the \citetalias{gontcharov2023} 3D map. The grid accounts for the variation in longitudinal step size with latitude, due to the influence of the cosine of $b$ on angular separation.
The grid in longitude is designed such that one grid point always lies at $l = 180\degr$, with all remaining grid points placed symmetrically around this central meridian. As a result, the grid avoids points near $l \approx 0\degr$, where large gradients in extinction and other observables toward the Galactic center make such locations less suitable for reliable averaging.

For each LOS in our 3D map, we adopt a uniform radial grid with a fixed step size of 50~pc, extending from the Sun out to the corresponding $R_{\mathrm{limit}}$.

While a uniform grid is convenient for map representation and for interpolating values at arbitrary spatial points, we adopt transverse and radial averaging windows in our 3D map that are not necessarily equal to the grid step. This approach ensures a sufficient number of stars per bin and accounts for the well-known correlation between adjacent LOSs \citep{green2019}.
Specifically, we adopt a constant transverse (angular) averaging window of 3.56~pc across the LOS for our 3D map --- an improvement over the approach in \citetalias{gontcharov2023}. This resolution is fine enough to capture spatial variations in extinction at the scale of individual dust clouds. Because the transverse window is fixed in linear size, its angular extent decreases with increasing distance --- from $4.07\degr$ at 50~pc to 6.1~arcminutes at 2~kpc. Thus, the averaging window matches the grid step only at $R = 2$~kpc; at smaller distances, the window spans a larger fraction of the sky relative to the grid.
In contrast to \citetalias{gontcharov2023}, who used a constant radial averaging window of 50~pc, we now account for the increasing {\it Gaia} parallax uncertainty with distance. Accordingly, we adopt a radial averaging window that increases linearly from 25~pc at $R = 50$~pc to 100~pc at $R = 2000$~pc.

To ensure high statistical precision in our results, we require that each 2D map cell or 3D map bin contains at least four stars, even in regions with sparse stellar density. If a given cell or bin contains fewer than four stars, we iteratively expand its transverse averaging window by a factor of 1.5 in each step until the required minimum is met.
As a result, only 1.8\% of the 2D map cells have an averaging window larger than 6.1 arcminutes. Similarly, fewer than 1\% of the 3D bins require a transverse averaging window larger than 3.56~pc. The final angular size of the averaging window adopted for each cell or bin is provided in the corresponding map tables.

\begin{figure} 
   \centering
   \includegraphics[width=14.0cm, angle=0]{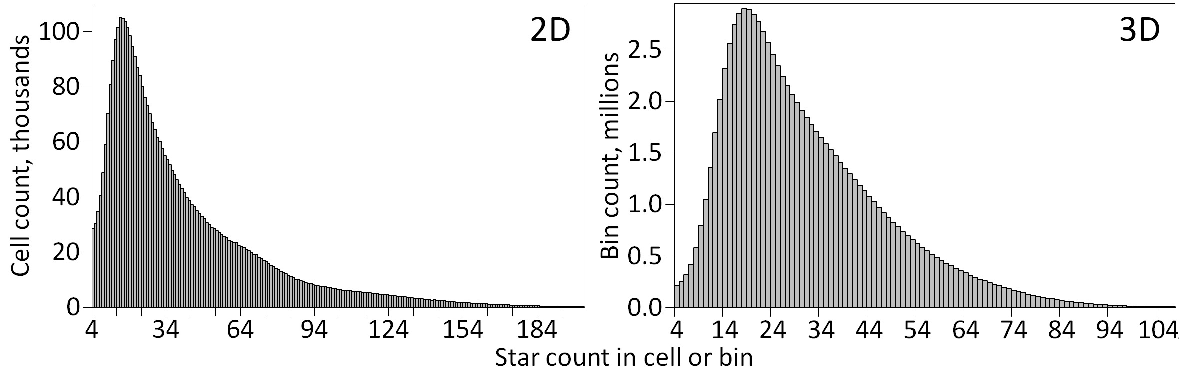}
\caption{Histograms showing the distribution of our 2D map cells and 3D map bins by the number of stars they contain. Cells and bins with higher star counts (up to 410 and 447 stars, respectively) are not displayed for clarity.} 
   \label{fig:histon.eps}
   \end{figure}

Fig.~\ref{fig:histon.eps} presents histograms of the number of stars per cell in our 2D map and per bin in our 3D map. The mode (median) of the star counts is 15 (29) for the 2D map cells and 18 (26) for the 3D map bins, respectively.

It is worth noting that the adopted 3D radial grid and averaging window --- 25 pc at best --- provide significantly lower resolution than the transverse resolution of 3.56~pc, primarily due to the relatively large uncertainties in distance $R$. More accurate parallaxes from future {\it Gaia} DR4, covering larger stellar samples, may enable the construction of 3D maps with substantially improved radial resolution.

There is the fundamental constraint imposed to 3D maps based on individual stellar reddening/extinction measurements. This constraint arises from the finite stellar density within the local part of the Galaxy. Following the Besan\c{c}on model of the Galaxy, the local spatial density of stars of all classes except stellar remnants (white dwarfs, neutron stars, and black holes of stellar mass) is about 0.04 solar mass per cubic parsec ($M_{\sun}pc^{-3}$) \citep{robin2022}. Most stars are M dwarfs of about 0.1 $M_{\sun}$ united in rather compact double or multiple systems with their typical mass of about 0.2 $M_{\sun}$. Each such system can be considered as a point object with one input reddening/extinction estimate for a 3D map. Hence, this provides a typical spatial density about one input estimate per 5 pc$^{3}$. On the other hand, given the aforementioned typical dust medium fluctuations about $\sigma(A_\mathrm{V})\approx0.3$\,mag at the Galactic mid-plane, one has to average about ten individual reddening/extinction estimates to achieve a rather high desired 3D map accuracy of about $\sigma(A_\mathrm{V})\approx0.1$\,mag. 
A typical volume containing ten stars (compact multiple stellar systems) is about 50 pc$^{3}$. This can be considered as a minimal bin of a 3D map with the same radial and transverse resolution. Taking into account that we have to omit peculiar stars, this resolution is 3--4 pc. As one moves away from the Galactic mid-plane, the spatial density of stars decreases, but the medium fluctuations decrease too. Therefore, the minimal bin is nearly the same, at least, within the Galactic dust layer, i.e. $|Z|<500$ pc. Thus, the transverse resolution of our 3D map is close to the minimal one, while the radial resolution should be improved by an order of magnitude in the future.

As in \citetalias{gontcharov2023}, the averaging of individual $A_\mathrm{V}$ values within spatial bins occasionally results in LOSs where the average extinction decreases with distance $R$. This non-physical behavior arises from fluctuations in the dust medium, uncertainties in individual $R$ and $A_\mathrm{V}$ estimates, as well as from a mismatch between the map's transverse resolution and dust cloud size \citep{gontcharov2017a}.
To suppress this effect, we iteratively adjust the average $A_\mathrm{V}$ values along each LOS by slightly increasing or decreasing adjacent values until a non-decreasing trend with $R$ is achieved. 
Namely, when the average $A_\mathrm{V}$ values are calculated for all bins of LOS, we fix the 2D map value to the one of the bin farthest w.r.t the Sun, and go from it to the bin nearest w.r.t. the Sun checking the non-decreasing of $A_\mathrm{V}$ with $R$ for each bin pair. For each inappropriate pair, we correct $A_\mathrm{V}$ in both the bins in such a way that the bin with lower $R$ becomes a small increment lower than the average $A_\mathrm{V}$ of the pair, while the bin with higher $R$ becomes the same increment higher than the average. Since a correction of a pair may lead to a correction of the next pair, some LOSs need iterations of this adjustment. Typically, this requires up to several dozen iterations. We found that the iterations may not converge if we adopt too large increment or, conversely, zero increment. Therefore, we empirically adopt the increment of $\Delta A_\mathrm{V}=0.0004$\,mag, which ensures the convergence for any LOS. Once started for all LOSs, this adjustment procedure automatically runs until complete.

Tables~\ref{table:2d} and \ref{table:3d} present our 2D and 3D map, respectively.
The 2D map has 3,991,111 cells in the sky, while our 3D map has 87,985,878 spatial bins for these cells.
The maps and tables from this paper are presented in Science Data Bank at \url{https://www.scidb.cn}.

\begin{figure} 
   \centering
   \includegraphics[width=14.0cm, angle=0]{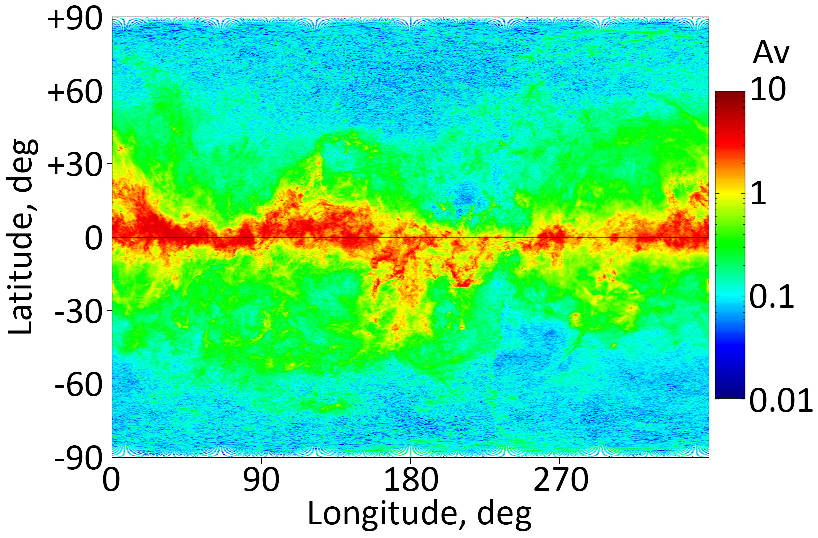}
\caption{Our 2D extinction map.}
   \label{fig:final2d.eps}
   \end{figure}

Fig.~\ref{fig:final2d.eps} presents our 2D extinction map. It closely resembles other widely used 2D maps of total Galactic extinction, reddening, or dust emission, such as those shown in figure 8 of \citetalias{sfd98}, figure 3 of \citetalias{gnilc}, figure 5 of \citetalias{green2019}, and figure 24 of \citetalias{dbb23}. While these maps share a similar overall structure, they differ primarily in their extinction estimates at high Galactic latitudes. A detailed comparison of these 2D maps is provided in Sect.~\ref{sect:2d}.

To estimate the accuracy of our extinction maps, we note that systematic uncertainties in $A_\mathrm{V}$ have been suppressed to the level of a few hundredths of a magnitude, as demonstrated in Sect.~\ref{sect:systematics}.
The statistical uncertainty can be estimated as follows: each cell or bin in our maps typically contains around 20 stars, with a minimum of 4 stars. Approximately 67\% of all cells and bins include more than 20 stars. For stars with $G < 18.5$~mag, \citetalias{anders2022} report a typical precision of $\sigma(A_\mathrm{V}) < 0.18$~mag for individual extinction estimates.
Therefore, the typical statistical uncertainty of our averaged $A_\mathrm{V}$ values is approximately $0.18 / \sqrt{20} \approx 0.04$~mag, and in the worst case (with 4 stars per bin), it does not exceed 0.09~mag.

As noted earlier, the uncertainty in extinction predictions for point sources arising from fluctuations in the interstellar dust medium ranges from approximately $\sigma(A_\mathrm{V}) = 0.06$~mag at high Galactic latitudes to $\sigma(A_\mathrm{V}) = 0.33$~mag or more near the Galactic plane. As a result, this source of uncertainty dominates the total error in $A_\mathrm{V}$ estimates for point objects across most of the sky, likely with the exception of the highest latitudes.
Therefore, in addition to their applicability for extended objects, our extinction maps (as well as any similar maps) are most appropriately used to estimate $A_\mathrm{V}$ for point sources that lack individual reddening or extinction estimates derived from their SEDs.\footnote{For example, within 2~kpc of the Sun, there are several billion stars without SED-based extinction estimates (accounting for about 99\% of all stars), including nearly 400 million stars from {\it Gaia} DR3 --- roughly 80\% of all {\it Gaia} DR3 stars in this volume.}

Fluctuations in the dust medium are less significant for extended celestial objects compared to point sources. To quantify their impact, we calculate the standard deviation of our $A_\mathrm{V}$ estimates across all LOSs intersecting each extended object considered further. For objects with known angular size, we use their full extent; for those without size information, we adopt a circular region with a radius of 6.1 arcminutes—the resolution of our 2D map.
This standard deviation is combined in quadrature with the statistical and systematic uncertainties to obtain the total uncertainty, $\sigma(A_\mathrm{V})$, for our extinction estimates.
It is important to note that the systematic uncertainty in both the 2D and 3D map predictions includes the effects of uncertainties in the parameters used for sample selection. In addition, the 3D map carries extra systematic uncertainty due to distance ($R$) errors. As a result, the total uncertainty in our extinction predictions for extended objects exceeds 0.07~mag for the 2D map and 0.10~mag for the 3D map.

\begin{table}
\bc
\begin{minipage}[]{100mm}
\def\baselinestretch{1}\normalsize\normalsize
\caption[]{Our 2D $A_\mathrm{V}$ map, fully available in electronic form.
}
\label{table:2d}
\end{minipage}
\setlength{\tabcolsep}{1pt}
\small
\begin{tabular}[c]{cccccc}
\hline\noalign{\smallskip}
$l$    &  $b$   & $R_{limit}$ & Window   & Number    & $A_\mathrm{V}$ \\
(degs) & (degs) & (parsecs)   & (degs) & of dwarfs & (mag)  \\
\hline\noalign{\smallskip}
180.00000 & $-90.00000$ &  450 & 0.102 &  10 & 0.097 \\
180.00000 & $-89.97500$ &  450 & 0.102 &  18 & 0.110 \\
180.00000 & $-89.87333$ &  450 & 0.102 &  16 & 0.102 \\
134.01410 & $-89.87333$ &  450 & 0.102 &   9 & 0.089 \\
 88.02814 & $-89.87333$ &  450 & 0.102 &   9 & 0.106 \\
\ldots    &   \ldots    & \ldots & \ldots & \ldots & \ldots \\
  \noalign{\smallskip}\hline
\end{tabular}
\ec
\end{table}

\begin{table}
\bc
\begin{minipage}[]{100mm}
\def\baselinestretch{1}\normalsize\normalsize
\caption[]{Our 3D $A_\mathrm{V}$ map, fully available in electronic form.
}
\label{table:3d}
\end{minipage}
\setlength{\tabcolsep}{1pt}
\small
\begin{tabular}[c]{cccccc}
\hline\noalign{\smallskip}
$l$       &     $b$     &  $R$ & Window   & Number    & $A_\mathrm{V}$  \\
(deg)     &    (deg)    & (pc) & (degs) & of dwarfs &   (mag)    \\
\hline\noalign{\smallskip}
  0.00000 & $-90.00000$ &  450 & 0.452 &  18 &  0.097 \\
  0.00000 & $-90.00000$ &  400 & 0.508 &  11 &  0.096 \\
  0.00000 & $-90.00000$ &  350 & 0.581 &   9 &  0.091 \\
  0.00000 & $-90.00000$ &  300 & 0.678 &  11 &  0.090 \\
  0.00000 & $-90.00000$ &  250 & 0.813 &  10 &  0.090 \\
  0.00000 & $-90.00000$ &  200 & 1.017 &  14 &  0.090 \\
  0.00000 & $-90.00000$ &  150 & 1.356 &  15 &  0.080 \\
  0.00000 & $-90.00000$ &  100 & 2.033 &  11 &  0.047 \\
  0.00000 & $-90.00000$ &   50 & 4.067 &  11 &  0.047 \\
\ldots    &   \ldots    & \ldots & \ldots & \ldots & \ldots \\
  \noalign{\smallskip}\hline
\end{tabular}
\ec
\end{table}

\begin{table}
\bc
\begin{minipage}[]{100mm}
\def\baselinestretch{1}\normalsize\normalsize
\caption[]{Our $A_\mathrm{V}$ estimates for galaxies and quasars, fully available in electronic form.
}
\label{table:galaxies}
\end{minipage}
\setlength{\tabcolsep}{1pt}
\small
\begin{tabular}[c]{cccccccc}
\hline\noalign{\smallskip}
Number & SIMBAD Name & $\alpha$ & $\delta$ &  $l$  &  $b$  & $A_\mathrm{V}$ & $\sigma(A_\mathrm{V})$ \\
       &             &   (deg)  &   (deg)  & (deg) & (deg) &     (mag)      &        (mag)    \\
\hline\noalign{\smallskip}
1 & [B68b] 142         & 194.3191 & $+36.7875$ & 116.0047 & $+80.2621$ & 0.09 & 0.07 \\
2 & [B68b] 194         & 194.6240 & $+35.4788$ & 113.1477 & $+81.5150$ & 0.11 & 0.07 \\
3 & [B68b] 201         & 194.9533 & $+34.3896$ & 109.5413 & $+82.5196$ & 0.10 & 0.07 \\
4 & [BIG2010] GNS-JD2  & 189.1061 & $+62.2421$ & 125.9636 & $+54.7977$ & 0.09 & 0.07 \\
5 & [BKG2010] 14       &  73.5518 &  $-3.0213$ & 201.5065 & $-27.3275$ & 0.17 & 0.07 \\
\ldots & \ldots & \ldots & \ldots & \ldots & \ldots & \ldots & \ldots \\
  \noalign{\smallskip}\hline
\end{tabular}
\ec
\tablecomments{0.86\textwidth}{The galaxies are sorted by their SIMBAD name. 
}
\end{table}

\begin{table}
\bc
\begin{minipage}[]{100mm}
\def\baselinestretch{1}\normalsize\normalsize
\caption[]{Our $A_\mathrm{V}$ estimates for SN~Ia, fully available in electronic form.
}
\label{table:snia}
\end{minipage}
\setlength{\tabcolsep}{1pt}
\small
\begin{tabular}[c]{ccccccccc}
\hline\noalign{\smallskip}
Number & SIMBAD Name & $\alpha$ & $\delta$ &  $l$  &  $b$  & Redshift & $A_\mathrm{V}$ & $\sigma(A_\mathrm{V})$ \\
       &             &   (deg)  &   (deg)  & (deg) & (deg) &          &   (mag)      &        (mag)    \\
\hline\noalign{\smallskip}
1 & [GBM2015] SDSS 1059-52618-553 SN & 117.3879 & $+27.9581$ & 192.6367 & $+24.2666$ & 0.12158 & 0.13 & 0.07 \\
2 & [GBM2015] SDSS 1167-52738-214 SN & 234.7342 & $+47.7628$ &  76.7587 & $+51.5414$ & 0.07001 & 0.08 & 0.07 \\
3 & [GBM2015] SDSS 1266-52709-24 SN  & 124.1958 & $+25.2919$ & 197.6424 & $+29.1820$ & 0.13976 & 0.14 & 0.07 \\
4 & [GBM2015] SDSS 1574-53476-461 SN & 245.8904 & $+25.4056$ &  43.4599 & $+42.7727$ & 0.19025 & 0.30 & 0.07 \\
5 & [GBM2015] SDSS 1605-53062-528 SN & 170.4500 & $+12.8806$ & 242.6984 & $+64.6518$ & 0.10110 & 0.09 & 0.07 \\
\ldots & \ldots & \ldots & \ldots & \ldots & \ldots & \ldots & \ldots & \ldots \\
  \noalign{\smallskip}\hline
\end{tabular}
\ec
\tablecomments{0.86\textwidth}{The SN~Ia are sorted by their SIMBAD name. 
}
\end{table}

\begin{table}
\bc
\begin{minipage}[]{100mm}
\def\baselinestretch{1}\normalsize\normalsize
\caption[]{Our $A_\mathrm{V}$ estimates for Galactic globular clusters, fully available in electronic form.
}
\label{table:gc}
\end{minipage}
\setlength{\tabcolsep}{1pt}
\small
\begin{tabular}[c]{ccccccccccccc}
\hline\noalign{\smallskip}
Number & SIMBAD Name & $\alpha$ & $\delta$ &  $l$  &  $b$  & Diameter & $R$  & $X$  & $Y$  & $Z$  & $A_\mathrm{V}$ & $\sigma(A_\mathrm{V})$ \\
       &             &   (deg)  &   (deg)  & (deg) & (deg) & (arcmin) & (pc) & (pc) & (pc) & (pc) &     (mag)      &        (mag)    \\
\hline\noalign{\smallskip}
1 & [FSR2007] 1716 & 342.6205 & $-53.7462$ & 329.7777 & $ -1.5870$ & 31.6 &   7431 &   $6419$ &  $-3739$ &   $-206$ & $>$ 1.90 & 0.84 \\
2 & 2MASS-GC03     & 253.0442 & $-47.0581$ & 339.1876 & $ -1.8532$ & 26.7 &   9082 &   $8485$ &  $-3225$ &   $-294$ & $>$ 3.16 & 0.14 \\
3 & 2MASS-GC01     & 272.0909 & $-19.8297$ &  10.4710 & $  0.1001$ & 62.1 &   3373 &   $3317$ &    $613$ &      $6$ & $>$ 2.70 & 0.19 \\
4 & 2MASS-GC02     & 272.4021 & $-20.7789$ &   9.7821 & $ -0.6152$ & 19.6 &   5503 &   $5423$ &    $935$ &    $-59$ & $>$ 2.17 & 0.29 \\
5 & NAME E 1       &  58.7600 & $-49.6067$ & 258.3487 & $-48.4728$ &  7.6 & 118905 & $-15920$ & $-77207$ & $-89017$ &    0.09 & 0.07 \\
\ldots & \ldots & \ldots & \ldots & \ldots & \ldots & \ldots & \ldots & \ldots & \ldots & \ldots & \ldots & \ldots \\
  \noalign{\smallskip}\hline
\end{tabular}
\ec
\tablecomments{0.86\textwidth}{The globular clusters are sorted by their SIMBAD name. 
The symbol `$>$' preceding the $A_\mathrm{V}$ value for objects with $|b| < 13\degr$ indicates that the estimate represents a lower limit on the true extinction.
}
\end{table}

\begin{table}
\bc
\begin{minipage}[]{100mm}
\def\baselinestretch{1}\normalsize\tiny
\caption[]{Our $A_\mathrm{V}$ estimates for Galactic open clusters, fully available in electronic form.
}
\label{table:open}
\end{minipage}
\setlength{\tabcolsep}{1pt}
\small
\begin{tabular}[c]{ccccccccccccccc}
\hline\noalign{\smallskip}
Number & SIMBAD Name & $\alpha$ & $\delta$ &  $l$  &  $b$  & Diameter & $\varpi$ & $R$  & $X$  & $Y$  & $Z$  & \citetalias{green2019} $A_\mathrm{V}$ & Our $A_\mathrm{V}$ & $\sigma(A_\mathrm{V})$ \\
  & &   (deg)  &   (deg)  & (deg) & (deg) & (arcmin) & mas & (pc) & (pc) & (pc) & (pc) & (mag) & (mag) &  (mag) \\
\hline\noalign{\smallskip}
1 & ESO 489-1         &  91.2417 & $-26.7350$ & 232.9298 & $-21.4198$ & 11.0 &  3.134 & 317 & $-178$ & $-236$ & $-116$ & 0.00 & 0.09 & 0.10 \\
2 & NGC 1662          &  72.1980 & $+10.8820$ & 187.7945 & $-21.0767$ & 18.0 &  2.400 & 413 & $-382$ & $ -52$ & $-149$ & 0.99 & 1.08 & 0.10 \\
3 & NGC 1333          &  52.2970 & $+31.3100$ & 158.3430 & $-20.5052$ & 22.0 &  3.344 & 297 & $-259$ & $ 103$ & $-104$ & 4.93 & 1.86 & 0.32 \\
4 & [KC2019] Theia 63 & 134.9705 & $-77.8048$ & 291.9390 & $-20.2281$ & 25.0 & 10.302 &  97 & $  34$ & $ -84$ & $ -33$ &      & 0.07 & 0.10 \\
5 & OCSN 70           &  85.2700 &  $-9.3800$ & 213.4862 & $-19.8568$ & 23.4 &  2.230 & 444 & $-349$ & $-231$ & $-151$ & 6.10 & 3.30 & 0.12 \\
\ldots & \ldots & \ldots & \ldots & \ldots & \ldots & \ldots & \ldots & \ldots & \ldots & \ldots & \ldots & \ldots & \ldots & \ldots \\
  \noalign{\smallskip}\hline
\end{tabular}
\ec
\tablecomments{0.86\textwidth}{The open clusters are sorted by ascending $b$.
$\varpi$ is the parallax from {\it Gaia} DR3.
$R=1000/(\varpi+0.02)$.
\citetalias{green2019} $A_\mathrm{V}$ is derived for $R$.
}
\end{table}

\begin{table}
\bc
\begin{minipage}[]{100mm}
\def\baselinestretch{1}\normalsize\normalsize
\caption[]{Our estimates for the molecular clouds from \citetalias{zucker2020}, fully available in electronic form.
}
\label{table:zss20}
\end{minipage}
\setlength{\tabcolsep}{1pt}
\small
\begin{tabular}[c]{ccccccccccc}
\hline\noalign{\smallskip}
Number & Name & $\alpha$ & $\delta$ &  $l$  &  $b$  & $R$  & $\sigma(R)$ & $A_\mathrm{V}^F$ & $A_\mathrm{V}^B$ & Note \\
       &      &   (deg)  &   (deg)  & (deg) & (deg) & (pc) &     (pc)    &     (mag)        &      (mag)       &  \\
\hline\noalign{\smallskip}
1 & Aquila Rift & 269.5 & $-5.6$ & 21.8 & $+9.2$  & 252 & 25 & 0.20 & 1.90 & \\	
2 & Aquila Rift	& 265.4 & $-9.0$ & 16.7 & $+11.1$ & 203 & 35 & 0.30 & 1.80 & \\
3 & Aquila Rift	& 264.8 & $-6.8$ & 18.3 & $+12.7$ & 262 & 30 & 0.22 & 2.86 & \\
4 & Aquila Rift	& 267.0 & $-4.5$ & 21.5 & $+11.9$ & 270 & 40 & 0.16 & 1.54 & \\
5 & Aquila Rift	& 260.3 & $-6.8$ & 16.0 & $+16.6$ & 172 & 35 & 0.09 & 2.25 & \\
\ldots & \ldots & \ldots & \ldots & \ldots & \ldots & \ldots & \ldots & \ldots & \ldots & \ldots \\
  \noalign{\smallskip}\hline
\end{tabular}
\ec
\tablecomments{0.86\textwidth}{We retain the numbering, names, sorting, and Galactic coordinates of the clouds from \citetalias{zucker2020}, and sort the list accordingly.
Cases where multiple clouds share the same LOS are marked with an asterisk in the `Note' column.}
\end{table}

\begin{table}
\bc
\begin{minipage}[]{100mm}
\def\baselinestretch{1}\normalsize\normalsize
\caption[]{Our estimates for the molecular clouds from \citetalias{chen2020}, fully available in electronic form.
}
\label{table:cly20}
\end{minipage}
\setlength{\tabcolsep}{1pt}
\small
\begin{tabular}[c]{cccccccccc}
\hline\noalign{\smallskip}
Number & $\alpha$ & $\delta$ &  $l$  &  $b$  & $R$  & $\sigma(R)$ & $A_\mathrm{V}^F$ & $A_\mathrm{V}^B$ & Note \\
       &   (deg)  &   (deg)  & (deg) & (deg) & (pc) &     (pc)    &     (mag)        &      (mag)       &  \\
\hline\noalign{\smallskip}
  1 &  76.026 & 25.460 & 177.727 & -9.596 & 187 & 35 & 0.12 & 2.00 & * \\
  2 &  74.652 & 27.016 & 175.715 & -9.651 & 192 & 35 & 0.11 & 1.65 & * \\
  3 &  71.319 & 31.749 & 170.131 & -8.972 & 197 & 25 & 0.18 & 2.40 &   \\
  4 & 348.800 & 50.740 & 107.778 & -9.278 & 180 & 25 & 0.08 & 0.27 & * \\
  5 & 332.356 & 44.556 &  94.961 & -9.339 & 471 & 25 & 0.37 & 0.92 &    \\
\ldots & \ldots & \ldots & \ldots & \ldots & \ldots & \ldots & \ldots & \ldots & \ldots \\
  \noalign{\smallskip}\hline
\end{tabular}
\ec
\tablecomments{0.86\textwidth}{We retain the numbering, names, sorting, and Galactic coordinates of the clouds from \citetalias{chen2020}, and sort the list accordingly.
Cases where multiple clouds share the same LOS are marked with an asterisk in the `Note' column.}
\end{table}

\section{Validation of Our Extinction Maps}           
\label{sect:tests}

To test the reliability of our maps, we compare their $A_\mathrm{V}$ predictions for selected samples of extended celestial objects with corresponding $A_\mathrm{V}$ estimates from other 2D and 3D extinction maps, as well as with independent values reported in the literature.

We aim to select extended objects with angular diameters between 2 and 40 arcminutes, as discussed in Sect.~\ref{sect:intro}. However, this criterion cannot always be strictly applied due to the large uncertainties in the angular sizes of some objects. As a result, our lists are not fully complete. In addition, we include several particularly interesting objects whose angular sizes slightly exceed 40 arcminutes.
We also compile a list of SN~Ia. Our extinction estimates for these objects may still be reasonably accurate and scientifically useful, as the host galaxies of at least some SN~Ia are extended and fall within the applicability range of our maps.

We compile samples of 19,809 galaxies and quasars (Table~\ref{table:galaxies}), 8,293 Type Ia supernovae (Table~\ref{table:snia}), 170 globular clusters (Table~\ref{table:gc}), and 458 open clusters (Table~\ref{table:open}), distributed across the entire sky. Among these, 18,087 galaxies and quasars, 8,138 SN~Ia, 73 globular clusters, and 6 open clusters are located behind the Galactic dust layer --- that is, at Galactic latitudes $|b| > 13\degr$.
In addition, we analyze our extinction prediction for dust/molecular clouds from two catalogs: 318 clouds from \citet[][hereafter ZSS20]{zucker2020} and 537 ones from \citet[][hereafter CLY20]{chen2020} (the remaining 8 and 30 clouds in the catalogs, respectively, appear too distant to be reliably detected in our maps).
Our results for the \citetalias{zucker2020} and \citetalias{chen2020} clouds are present in Tables~\ref{table:zss20} and \ref{table:cly20}, respectively.

In Tables~\ref{table:galaxies}, \ref{table:snia}, and \ref{table:gc}, we place the symbol `$>$' before our $A_\mathrm{V}$ estimates for objects located at Galactic latitudes $|b| < 13\degr$. These estimates should be interpreted as lower limits to the true extinction and used with caution, as the objects may lie within the Galactic dust layer or beyond the reliable distance range of our extinction maps.

\begin{figure} 
   \centering
   \includegraphics[width=14.0cm, angle=0]{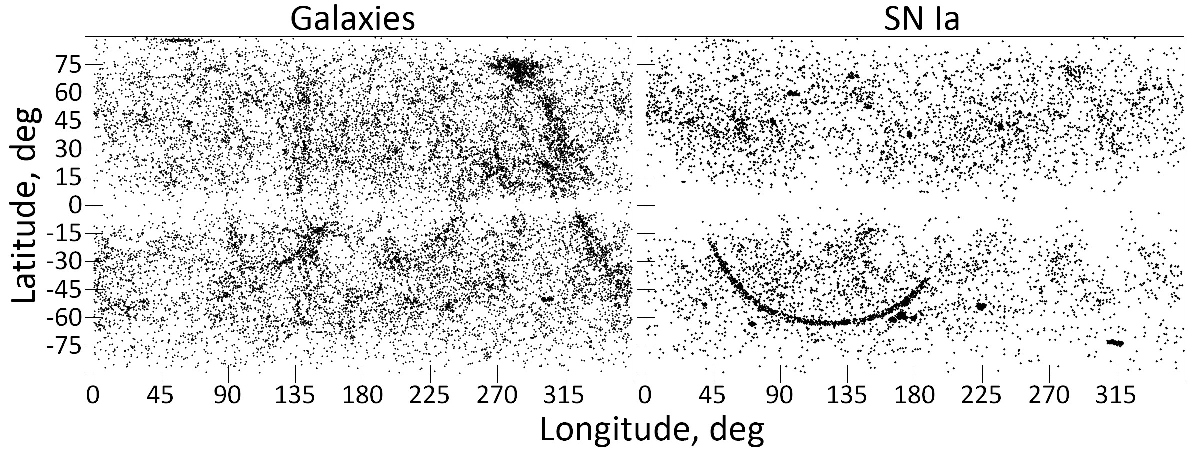}
   \caption{Sky distribution of the selected galaxies and SN~Ia in Galactic coordinates.}
   \label{fig:galaxies_sn_lb.eps}
   \end{figure}

\subsection{Galaxies and SN~Ia}           
\label{sect:galaxies}

The list of galaxies and quasars was compiled using the SIMBAD astronomical database \citep{simbad}\footnote{\url{https://simbad.cds.unistra.fr/simbad}}, the HyperLeda database \citep{hyperleda}\footnote{\url{http://leda.univ-lyon1.fr/}}, and the NASA/IPAC Extragalactic Database (NED) \citep{ned}\footnote{\url{https://ned.ipac.caltech.edu}}. For some objects, the angular size in the optical range is limited, uncertain, or even inconsistent across sources. Therefore, we retain in our list all galaxies and quasars for which a diameter in the range of 2--40 arcminutes in the optical band can be reasonably suggested.

The list of supernovae was compiled using the SIMBAD astronomical database.

Fig.~\ref{fig:galaxies_sn_lb.eps} shows the sky distribution of the selected galaxies and SN~Ia in Galactic coordinates, which appears as expected.

\subsection{Globular clusters}           
\label{sect:globulars}

The list of Galactic globular clusters was compiled using the SIMBAD astronomical database and the catalog of \citet{bica2019}. We exclude globular clusters associated with the Magellanic Clouds. Cluster distances are primarily adopted from \citet{baumgardt2021}, with the exception of distances for GLIMPSEC01 \citep{hare2018}, GLIMPSEC02 \citep{davidge2016}, and Gran2, Gran5, Patchick126, and VVV~CL160 \citep{bica2024}.

\begin{figure} 
   \centering
   \includegraphics[width=14.0cm, angle=0]{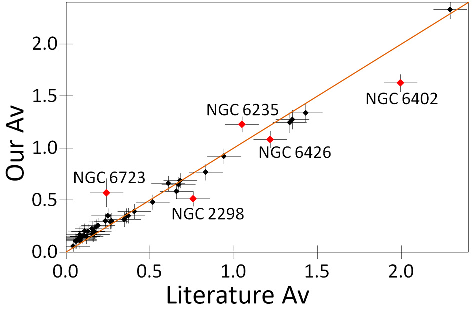}
   \caption{Comparison of our $A_\mathrm{V}$ estimates with literature values for 45 Galactic globular clusters. The orange line indicates the one-to-one correspondence. Outliers are highlighted in red.}
   \label{fig:gc.eps}
   \end{figure}

\begin{figure} 
   \centering
   \includegraphics[width=14.0cm, angle=0]{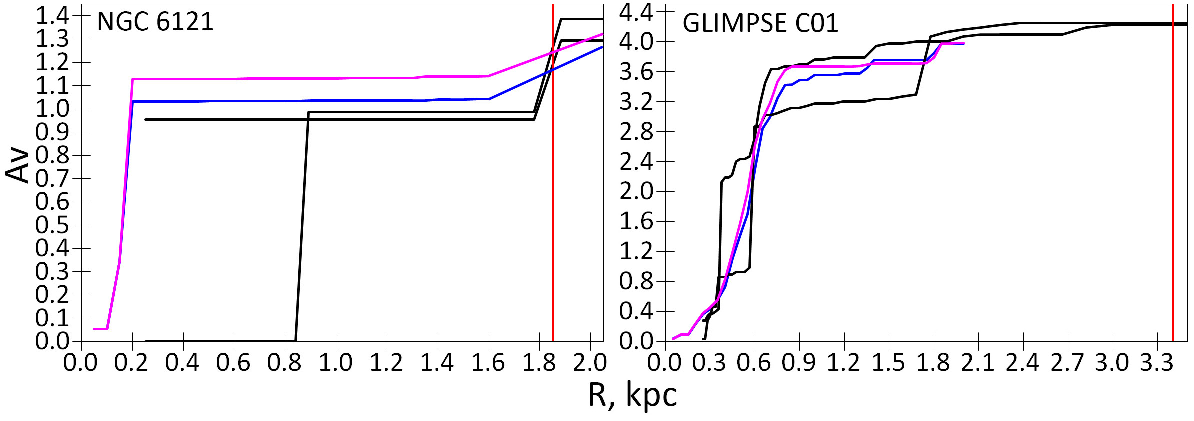}
\caption{Extinction as a function of distance along selected LOSs toward the globular clusters NGC\,6121 and GLIMPSEC01, based on our estimates (blue and magenta lines) and those from \citetalias{green2019} (black lines). For clarity, only two LOSs are shown for each map. The red lines indicate the assumed cluster distances, adopted from \citet{baumgardt2021} for NGC\,6121 and \citet{hare2018} for GLIMPSEC01.}
   \label{fig:gc_ngc6121_glimpse_c01.eps}
   \end{figure}

To validate the $A_\mathrm{V}$ estimates from our 2D map, we compare them with independent extinction estimates for Galactic globular clusters located at $|b| > 13\degr$. These independent values include estimates from \citet{clementini2023}, who analyzed RR~Lyrae variables in the clusters NGC\,288, NGC\,5139, and IC\,4499, as well as estimates from various authors based on photometric fitting to theoretical isochrones in CMDs and other methods \citep{recio2005,cassisi2008,dotter2011,koch2014,ngc5904,ngc6205,ngc288,yepez2022,ngc6362,ngc6397,ngc5024}.\footnote{This includes our own CMD-based estimates, which are fully independent of any reddening/extinction map.}

We adopt a conservative uncertainty of 0.1~mag for the literature $A_\mathrm{V}$ values. We restrict the comparison to 45 clusters located within 25~kpc of the Sun, using distances from \citet{baumgardt2021}, as literature-based extinction estimates become significantly less reliable at greater distances.

Fig.~\ref{fig:gc.eps} shows good overall agreement between our 2D map estimates and those from the literature, with the exception of five outliers, highlighted in red. The discrepancies for these outliers can likely be attributed to strong differential reddening (i.e., steep reddening gradients) in their surrounding regions, as previously reported by \citet{legnardi2023} for NGC\,2298, \citet{alonso2012} for NGC\,6235, \citet{yepez2022} for NGC\,6402, \citet{bonatto2013} for NGC\,6426, and by us for NGC\,6723 likely due to the nearby Corona Australis molecular cloud complex  \citep{ngc6362}.
A comparison of Fig.~\ref{fig:gc.eps} with figure 6 from \citetalias{gontcharov2023} demonstrates an improved agreement between our updated extinction estimates and those from the literature.

Fig.~\ref{fig:gc.eps} shows that our $A_\mathrm{V}$ estimates tend to be slightly higher than literature values at low extinction and slightly lower at high extinction, although the differences remain within the stated uncertainties. This trend may be attributed to spatial variations in extinction law or to an imperfect treatment of the law.
For example, to convert reddening estimates from \citet{dotter2010} and \citet{dotter2011} into $A_\mathrm{V}$, we adopt extinction coefficients of $A_\mathrm{V}/E(F606W{-}F814W) = 3.19$ and $A_\mathrm{V}/E(B{-}V) = 3.38$, based on the calibrations of \citet{casagrande2014,casagrande2018a} for a typical effective temperature of $T_\mathrm{eff} = 5400$~K, characteristic of globular cluster members. However, these coefficients depend on stellar color or $T_\mathrm{eff}$, which in turn depend on cluster age and metallicity. Uncertainties in these parameters contribute to the residual discrepancies between our extinction estimates and those reported in the literature.

Some globular clusters are sufficiently nearby to be matched with our 3D extinction map. Two such examples are presented in Fig.~\ref{fig:gc_ngc6121_glimpse_c01.eps}, which shows extinction profiles (i.e., the variation of $A_\mathrm{V}$ with distance) from our 3D map and from \citetalias{green2019} along two representative LOSs in the direction of each cluster.
Our 3D map demonstrates a high sensitivity to foreground extinction variations and generally agrees with the \citetalias{green2019} profiles --- particularly near the distances of the clusters, which is a key region for validation. In the case of NGC\,6121, our map clearly identifies a foreground dust cloud at a well-defined distance of $R \approx 175 \pm 25$~pc. In contrast, \citetalias{green2019} suggests either $R \approx 860$~pc or $R < 250$~pc for the same feature, depending on the LOS, reflecting greater uncertainty.
This discrepancy likely arises because \citetalias{green2019} does not provide extinction estimates for $R < 250$~pc, where the dust cloud is likely located. The ability of our map to resolve extinction structure at distances below 250 pc --- unlike \citetalias{green2019} --- is a clear advantage of our approach and one of the primary motivations for its development.

\begin{figure} 
   \centering
   \includegraphics[width=14.0cm, angle=0]{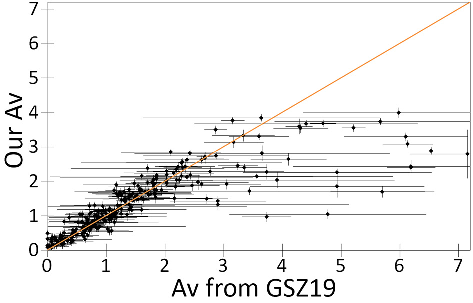}
\caption{Comparison of our $A_\mathrm{V}$ estimates with those from \citetalias{green2019} for 293 open clusters. The orange line indicates the one-to-one correspondence.}
\label{fig:oc_gsz19_gms25.eps}
   \end{figure}

\begin{figure} 
   \centering
   \includegraphics[width=14.0cm, angle=0]{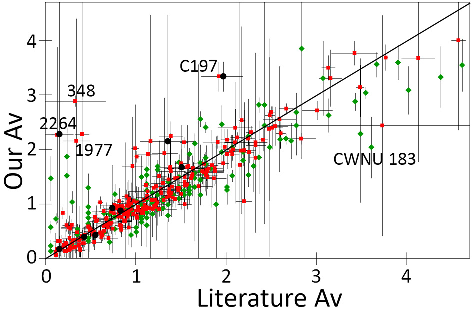}
\caption{Comparison of our $A_\mathrm{V}$ estimates with literature values for open clusters: 250 clusters from \citet{dias2021} (red squares), 129 clusters from \citet{he2022} (green diamonds), and 9 clusters from \citet{jackson2022} (black circles). The black line indicates the one-to-one correspondence. Notable outliers are labeled as ‘2264’ (NGC\,2264), ‘C197’ (Collinder\,197), ‘348’ (IC\,348), ‘1977’ (NGC\,1977), and CWNU\,183.}
   \label{fig:oc.eps}
   \end{figure}

\begin{figure} 
   \centering
   \includegraphics[width=14.0cm, angle=0]{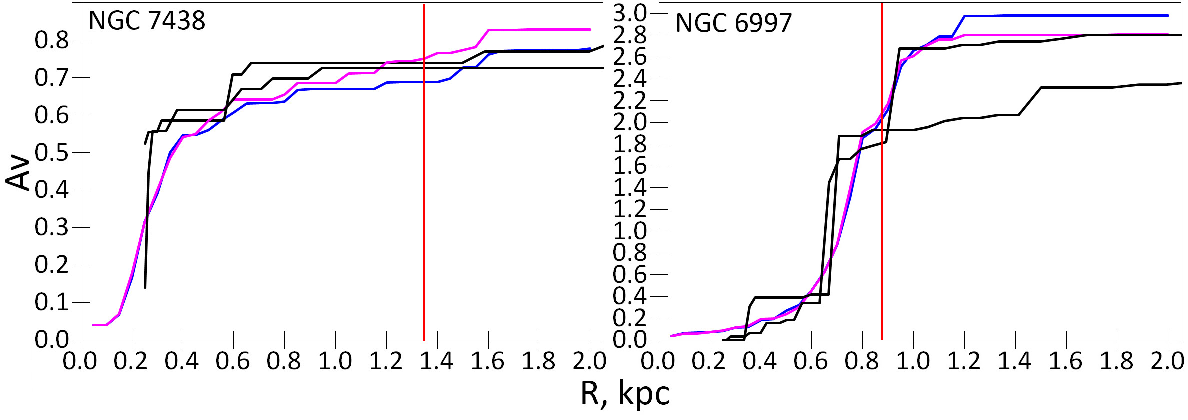}
\caption{Extinction as a function of distance along selected LOSs toward the open clusters NGC\,7438 and NGC\,6997, based on our estimates (blue and magenta lines) and those from \citetalias{green2019} (black lines). For clarity, only two LOSs are shown for each map. The red line indicates the cluster distance derived from {\it Gaia} DR3.}
   \label{fig:oc_ngc7438_ngc6997.eps}
   \end{figure}

\subsection{Open clusters}           
\label{sect:oc}

The list of open clusters was compiled using the SIMBAD astronomical database and the catalog of \citet{bica2019}. Open clusters are typically located within the Galactic dust layer. Therefore, to estimate their $A_\mathrm{V}$ values, we rely on our 3D extinction map: we construct extinction profiles along LOSs in the vicinity of each cluster and extract the $A_\mathrm{V}$ value at the cluster's distance.
Cluster distances are derived from {\it Gaia} DR3 parallaxes, incorporating the typical parallax zero-point correction of $+0.02$~mas, as recommended by \citet{lindegren2021}. This approach ensures a sufficient level of accuracy. We apply a uniform correction to the parallax values rather than correcting individual cluster members, as membership lists are often incomplete and uncertain.
The requirement for reliable {\it Gaia} DR3 parallaxes significantly reduces the number of usable open clusters in our sample. In addition, we exclude clusters associated with the Magellanic Clouds. Ultimately, we select 458 open clusters with corrected parallaxes greater than 0.5~mas --- that is, located approximately within 2~kpc of the Sun --- with one notable exception: the interesting cluster NGC\,2420, located at $R \approx 2611$~pc.

Among the selected open clusters, 293 have extinction estimates available from both \citetalias{green2019} and our study. A comparison of the two sets of estimates is shown in Fig.~\ref{fig:oc_gsz19_gms25.eps}. We find good agreement for 255 clusters (87\%) with $A_\mathrm{V} < 2.7$~mag. However, for clusters with higher extinction, significant discrepancies emerge.
These differences may be attributed to the lower angular resolution of our 3D map compared to that of \citetalias{green2019} (6.1 vs. 3.4 arcminutes), which allows the latter to resolve small-scale high-extinction structures more effectively. Additionally, our method may favor the selection of stars with lower extinction and be more affected by the obscuration of high-extinction stars, potentially biasing the derived $A_\mathrm{V}$ values downward relative to \citetalias{green2019}.

A comparison of our $A_\mathrm{V}$ estimates with literature values for 250 open clusters from \citet{dias2021}, 129 clusters from \citet{he2022}, and 9 clusters from \citet{jackson2022} is shown in Fig.~\ref{fig:oc.eps}. Several outliers are labeled. They are well-known young clusters embedded in interstellar clouds, where steep extinction gradients are present. Such gradients are partially smoothed in our map due to its finite resolution, which likely contributes to the discrepancies.

The total uncertainty in our $A_\mathrm{V}$ estimates for open clusters includes statistical and systematic components, as well as contributions from the parallax uncertainty (adopted as 0.01~mas from \citealt{vasiliev2021}) and uncertainties associated with the extinction law and its spatial or spectral variations.
The latter is particularly important. For example, we adopt a ratio of the extinction $A_\mathrm{V}$ to the reddening $E(BP{-}RP)$ between the {\it Gaia} filters as $A_\mathrm{V}/E(BP{-}RP) = 2.2$, based on \citet{casagrande2014,casagrande2018a}, assuming a typical effective temperature of $T_\mathrm{eff} = 7000$ K for cluster members. However, this coefficient is highly sensitive to stellar color and $T_\mathrm{eff}$. Consequently, the intrinsic spread in effective temperatures among cluster members introduces uncertainty when converting the initially derived reddening $E(BP{-}RP)$ into $A_\mathrm{V}$.
This effect may explain systematic differences between our $A_\mathrm{V}$ estimates and those of other studies. For instance, in Fig.~\ref{fig:oc.eps}, the green diamonds representing the results from \citet{he2022} tend to fall below the one-to-one relation, suggesting a systematic offset possibly driven by differences in the adopted extinction coefficients.

Fig.~\ref{fig:oc_ngc7438_ngc6997.eps} presents two examples of extinction profiles from our 3D map and from \citetalias{green2019} along selected LOSs in the regions of the open clusters NGC\,7438 and NGC\,6997. As with the globular clusters, our extinction profiles generally agree with those from \citetalias{green2019}, but our 3D map provides $A_\mathrm{V}$ estimates within $R < 250$~pc, where the \citetalias{green2019} map becomes uncertain. Notably, our map identifies a foreground dust cloud in front of NGC\,7438 at $R \approx 210 \pm 25$~pc.
In the case of NGC\,6997, the sharp rise in extinction profiles near the adopted cluster distance supports its classification as an embedded cluster within the North America Nebula. Interestingly, both our map and \citetalias{green2019} reveal a double-peaked structure in the extinction profiles, with the rises at $R \approx 700$ and $870$~pc, suggesting that NGC\,6997 may lie between two dense layers of the cloud. Additionally, the large discrepancy between the \citetalias{green2019} profiles at $R > 900$~pc indicates substantial internal extinction gradients within the cloud and possibly within the cluster itself.
Our 3D extinction map may thus serve as a valuable tool for future detailed studies of open cluster properties, including age determination --- since embedded clusters are typically younger than those that have already dispersed their natal gas and dust.

\subsection{Clouds}           
\label{sect:clouds}

Numerous catalogs of interstellar clouds within 2~kpc of the Sun exist, but they often differ in naming conventions, spatial boundaries, and hierarchical structure. For instance, the same cloud may have inconsistent coordinates in SIMBAD and in the catalog of \citetalias{zucker2020}. Such inconsistencies complicate direct comparisons between studies and hinder the synthesis of results. Consequently, a dedicated and systematic study of interstellar clouds is necessary to resolve these discrepancies and establish a unified framework.

Here, we test the performance of our 3D map by evaluating its ability to determine the distance $R$, foreground extinction $A_\mathrm{V}^F$, and backside extinction $A_\mathrm{V}^B$ for 318 and 537 molecular clouds from the recent catalogs by \citetalias{zucker2020} and \citetalias{chen2020}, respectively.
The cloud distance is defined as the location along the LOS where the steepest rise in the extinction profile is observed.
In cases where multiple such rises are present, we adopt the first among the rather steep ones --- corresponding to the noticeable cloud closest to the Sun along the LOS. 
Most LOSs from both \citetalias{zucker2020} and \citetalias{chen2020} --- particularly at low Galactic latitudes --- intersect multiple clouds. Such cases are indicated with an asterisk in Tables~\ref{table:zss20} and \ref{table:cly20}.
Since such rises along the same LOS are of different height, the selection of the desired rise is somewhat arbitrary. This requires a future detailed cloud-by-cloud study.

The lower and upper bounds of the steepest rise are used to estimate $A_\mathrm{V}^F$ and $A_\mathrm{V}^B$, respectively. Also, the distances of these bounds affect the derived distance total uncertainty $\sigma(R)$, which includes all statistical and systematic contributions.

\begin{figure} 
   \centering
   \includegraphics[width=14.0cm, angle=0]{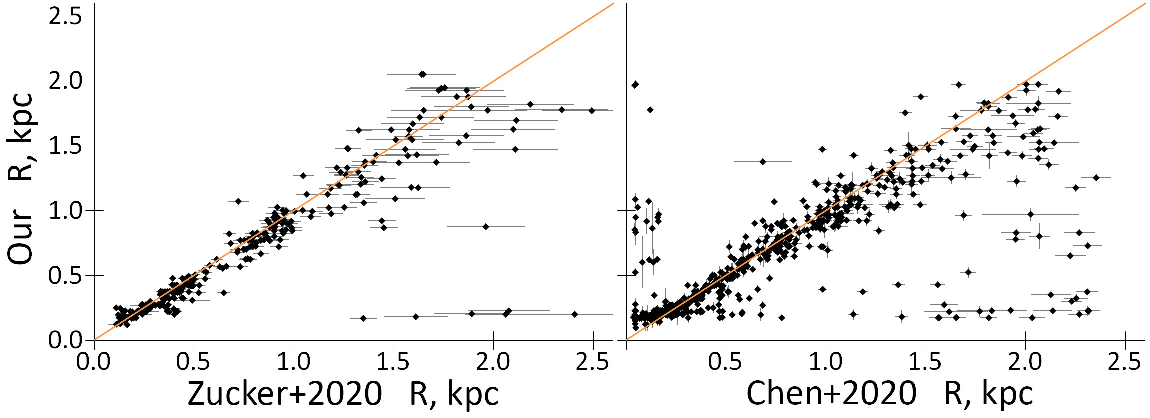}
\caption{Comparison of our estimated distances $R$ for molecular clouds with those from \citetalias{zucker2020} and \citetalias{chen2020}.
The orange line represents the one-to-one correspondence.}
   \label{fig:clouds_rr.eps}
   \end{figure}

\begin{figure} 
   \centering
   \includegraphics[width=14.0cm, angle=0]{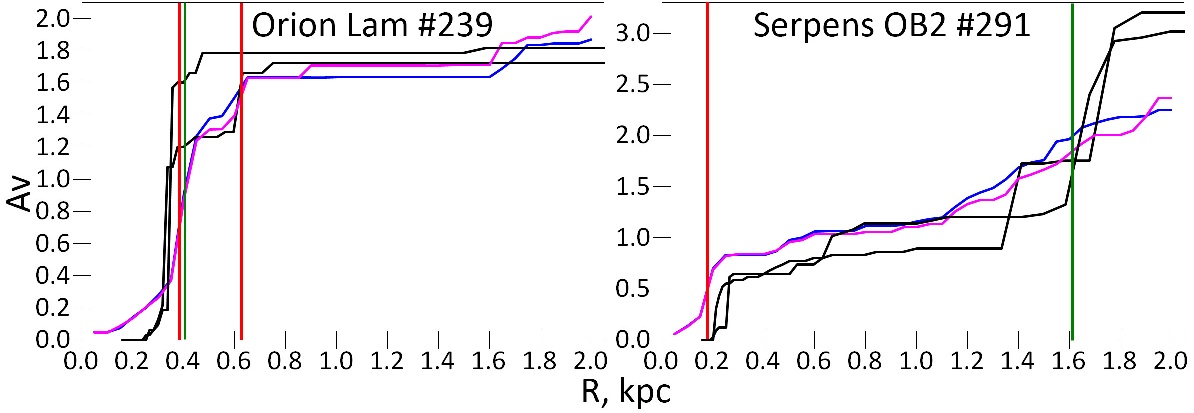}
   \caption{Extinction as a function of distance for selected LOSs toward \citetalias{zucker2020} clouds \#239 (OrionLam) and \#291 (SerpensOB2), based on our estimates (blue and magenta lines) and those from \citetalias{green2019} (black lines).
For clarity, only two LOSs are shown for each map.
For cloud \#239, the green line marks the distance $R = 406^{+20}_{-20}$~pc reported by \citetalias{zucker2020}, while the red lines indicate two clouds detected along these LOSs in our map at $R = 375 \pm 25$ and $625 \pm 25$~pc.
For cloud \#291, the green line corresponds to the distance $R = 1611 \pm 161$~pc from \citetalias{zucker2020}, and the red lines show a foreground cloud detected by us at $R = 175 \pm 25$~pc.
   }
   \label{fig:cloud_orion_lam239_serpensob2_291.eps}
   \end{figure}

\begin{figure} 
   \centering
   \includegraphics[width=14.0cm, angle=0]{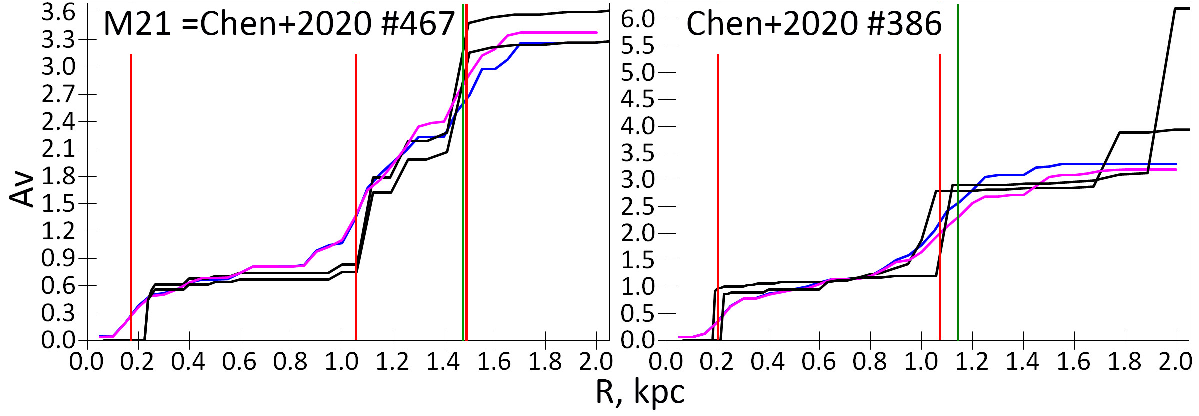}
   \caption{Extinction as a function of distance for selected LOSs toward \citetalias{chen2020} clouds \#467 (Messier~21) and \#386, based on our estimates (blue and magenta lines) and those from \citetalias{green2019} (black lines).
For clarity, only two LOSs are shown for each map.
For cloud \#467, the green line marks the distance $R = 1472\pm35$~pc reported by \citetalias{chen2020}, while the red lines indicate three clouds detected along these LOSs in our map at $R = 174 \pm 25$, $1054 \pm 25$, and $1477 \pm 25$~pc.
For cloud \#386, the green line corresponds to the distance $R = 1136 \pm 27$~pc from \citetalias{chen2020}, and the red lines show two clouds detected by us at $R = 200 \pm 25$ and $1073 \pm 35$~pc.
   }
   \label{fig:cloudschen.eps}
   \end{figure}

\begin{figure} 
   \centering
   \includegraphics[width=14.0cm, angle=0]{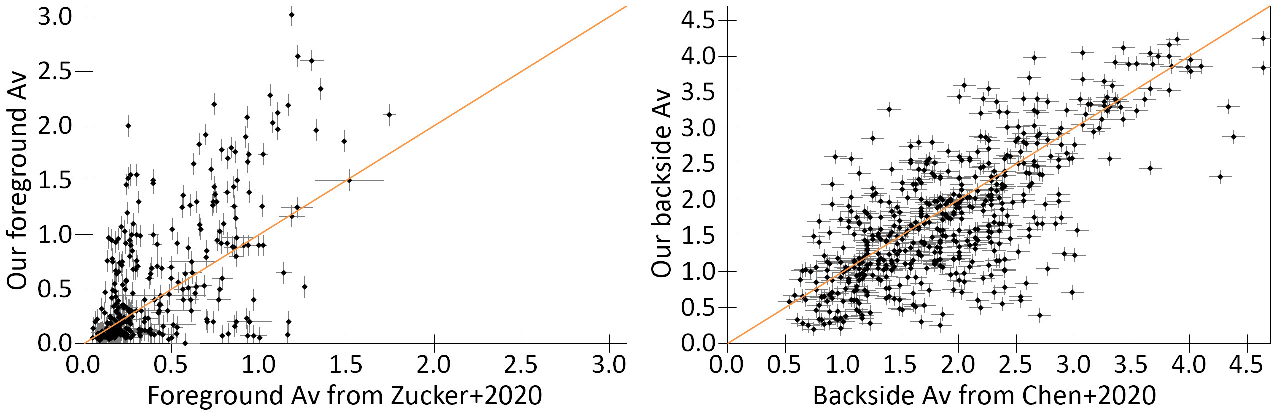}
\caption{Comparison of our foreground extinction estimates $A_\mathrm{V}^F$ for molecular clouds with those from \citetalias{zucker2020} and our backside extinction estimates $A_\mathrm{V}^B$ for molecular clouds with those from \citetalias{chen2020}.
The orange line indicates the one-to-one relation.
We adopt uncertainty $\sigma(A_\mathrm{V})=0.1$\,mag for our estimates and those from \citetalias{chen2020}.}
   \label{fig:clouds_av.eps}
   \end{figure}

Our $R$ estimates for the clouds are compared with those from \citetalias{zucker2020} and \citetalias{chen2020} in Fig.~\ref{fig:clouds_rr.eps}. 
Overall, the distances agree well. A slight systematic trend may exist in the sense that our $R$ is higher for nearby and lower for distant clouds w.r.t. those from both \citetalias{zucker2020} and \citetalias{chen2020}. This trend may be due to the fact that both \citetalias{zucker2020} and \citetalias{chen2020} use the {\it Gaia} DR2 parallaxes, while we use those from {\it Gaia} DR3.
Significant discrepancies between the distance estimates occur mainly in cases where multiple clouds lie along the same LOS and different clouds have been selected for comparison.
Such examples are presented in Figs~\ref{fig:cloud_orion_lam239_serpensob2_291.eps} and \ref{fig:cloudschen.eps}: sometimes we detect a nearby cloud, which is not detected by \citetalias{zucker2020} or \citetalias{chen2020} or, vice versa, we do not detect a nearby cloud detected by one of them.

Our estimates of $A_\mathrm{V}^F$ and $A_\mathrm{V}^B$ are compared in Fig.~\ref{fig:clouds_av.eps} with those of  \citetalias{zucker2020} and \citetalias{chen2020}, respectively (\citetalias{zucker2020} and \citetalias{chen2020} provide no backside and foreground extinction estimates, respectively). The \citetalias{chen2020} backside extinction $A_\mathrm{V}$ is calculated from $E(BP{-}RP)$ with $A_\mathrm{V}/E(BP{-}RP)=2.33$, both taken from \citetalias{chen2020}.
The presence of multiple clouds along the same LOS is likely a key factor contributing to the discrepancies between our $A_\mathrm{V}^F$ and $A_\mathrm{V}^B$ estimates and those from \citetalias{zucker2020} and \citetalias{chen2020} in Fig.~\ref{fig:clouds_av.eps}. A better agreement for the backside extinction suggests it is defined better than the foreground one.
Anyway, the determination of foreground or backside extinction to interstellar clouds seems to be more complex than often assumed and warrants further detailed investigation.

\subsection{Comparison of 2D maps}
\label{sect:2d}

\begin{figure} 
   \centering
   \includegraphics[width=14.0cm, angle=0]{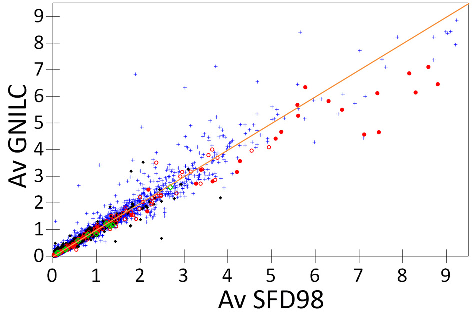}
\caption{Comparison of $A_\mathrm{V}$ from \citetalias{sfd98} and \citetalias{gnilc} for various object types: galaxies and quasars (blue crosses), SN~Ia (black diamonds), open clusters located behind the Galactic dust layer (open green diamonds), Galactic globular clusters behind the layer (open red circles), and those within the layer (filled red circles).
The orange line represents the one-to-one relation.}
   \label{fig:sfd98_gnilc.eps}
   \end{figure}

\begin{figure} 
   \centering
   \includegraphics[width=14.0cm, angle=0]{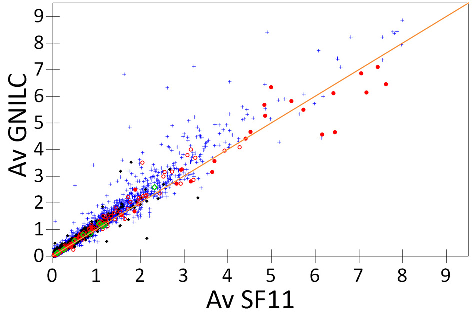}
   \caption{The same as Fig.~\ref{fig:sfd98_gnilc.eps} but for \citetalias{schlaflyfinkbeiner2011} vs \citetalias{gnilc}.}
   \label{fig:sf11_gnilc.eps}
   \end{figure}
   
\begin{figure} 
   \centering
   \includegraphics[width=14.0cm, angle=0]{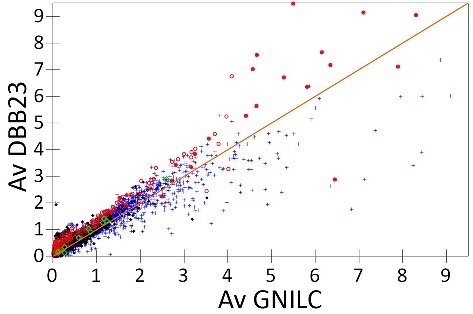}
   \caption{The same as Fig.~\ref{fig:sfd98_gnilc.eps} but for \citetalias{gnilc} vs \citetalias{dbb23}.}
   \label{fig:gnilc_dbb23.eps}
   \end{figure}

\begin{figure} 
   \centering
   \includegraphics[width=14.0cm, angle=0]{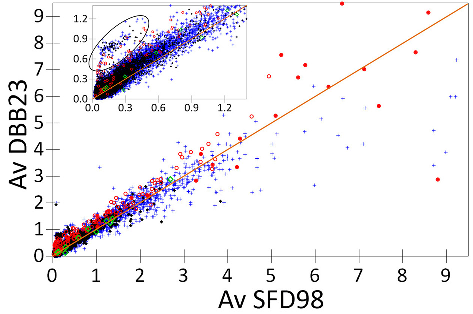}
   \caption{The same as Fig.~\ref{fig:sfd98_gnilc.eps} but for \citetalias{sfd98} vs \citetalias{dbb23}.
   The inset enlarges the low extinction domain.}
   \label{fig:sfd98_dbb23.eps}
   \end{figure}

\begin{figure} 
   \centering
   \includegraphics[width=14.0cm, angle=0]{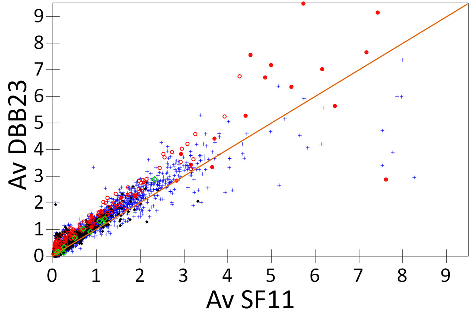}
   \caption{The same as Fig.~\ref{fig:sfd98_gnilc.eps} but for \citetalias{schlaflyfinkbeiner2011} vs \citetalias{dbb23}.}
   \label{fig:sf11_dbb23.eps}
   \end{figure}

\begin{figure} 
   \centering
   \includegraphics[width=14.0cm, angle=0]{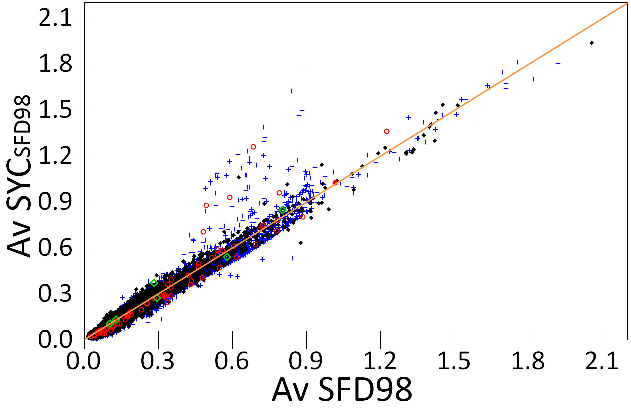}
   \caption{The same as Fig.~\ref{fig:sfd98_gnilc.eps} but for \citetalias{sfd98} vs SYC$_\mathrm{SFD98}$.}
   \label{fig:sfd98_syc.eps}
   \end{figure}
   
\begin{figure} 
   \centering
   \includegraphics[width=14.0cm, angle=0]{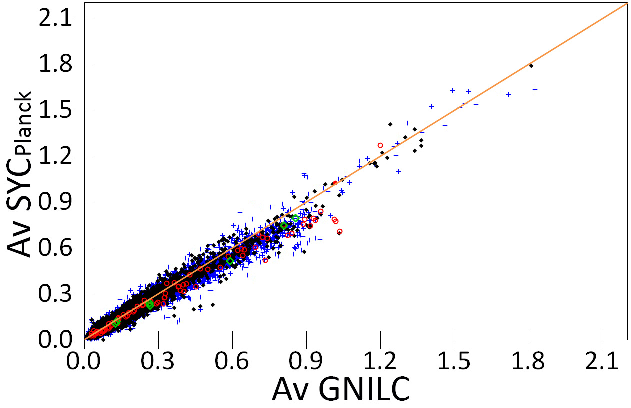}
   \caption{The same as Fig.~\ref{fig:sfd98_gnilc.eps} but for \citetalias{gnilc} vs SYC$_\mathrm{Planck}$.}
   \label{fig:gnilc_syc.eps}
   \end{figure}

\begin{figure} 
   \centering
   \includegraphics[width=14.0cm, angle=0]{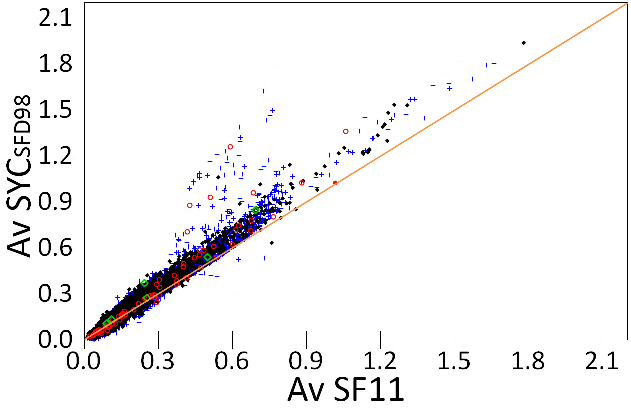}
   \caption{The same as Fig.~\ref{fig:sfd98_gnilc.eps} but for \citetalias{schlaflyfinkbeiner2011} vs SYC$_\mathrm{SFD98}$.}
   \label{fig:sf11_sycsfd98.eps}
   \end{figure}

\begin{figure} 
   \centering
   \includegraphics[width=14.0cm, angle=0]{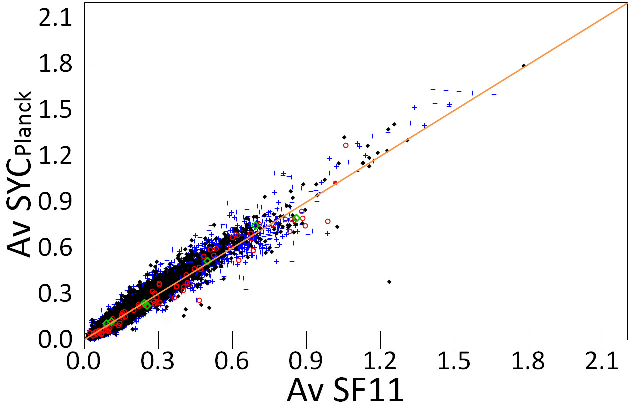}
   \caption{The same as Fig.~\ref{fig:sfd98_gnilc.eps} but for \citetalias{schlaflyfinkbeiner2011} vs SYC$_\mathrm{Planck}$.}
   \label{fig:sf11_sycplanck.eps}
   \end{figure}

\begin{figure} 
   \centering
   \includegraphics[width=14.0cm, angle=0]{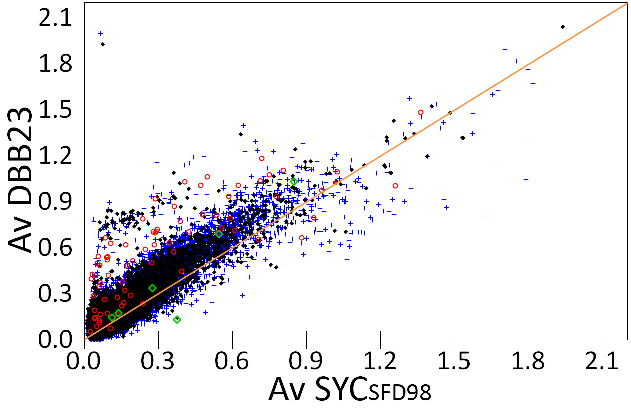}
   \caption{The same as Fig.~\ref{fig:sfd98_gnilc.eps} but for SYC$_\mathrm{SFD98}$ vs \citetalias{dbb23}.}
   \label{fig:sycsfd98_dbb23.eps}
   \end{figure}

\begin{figure} 
   \centering
   \includegraphics[width=14.0cm, angle=0]{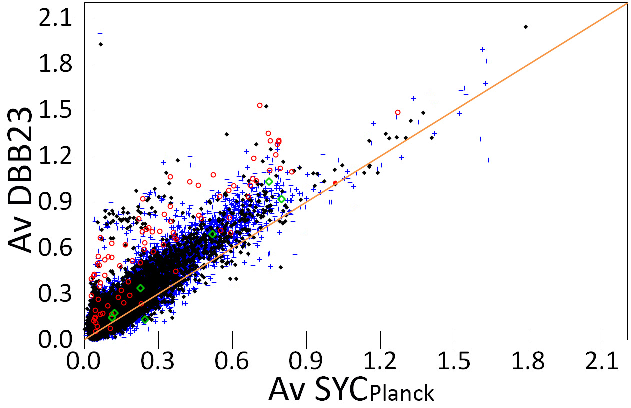}
   \caption{The same as Fig.~\ref{fig:sfd98_gnilc.eps} but for SYC$_\mathrm{Planck}$ vs \citetalias{dbb23}.}
   \label{fig:sycplanck_dbb23.eps}
   \end{figure}

We compare our 2D extinction map with the widely used 2D maps by \citetalias{dbb23}, \citetalias{sfd98}, \citetalias{schlaflyfinkbeiner2011}, \citetalias{gnilc}, and \citetalias{chiang2023}. Also, we consider recent recalibrations of the \citetalias{sfd98} and {\it Planck} \citep{planck2019} estimates in the high and middle Galactic latitudes by \citet{sun2022}, which are hereafter referred to as SYC$_\mathrm{SFD98}$ and SYC$_\mathrm{Planck}$, respectively. Note that most SYC$_\mathrm{SFD98}$ and SYC$_\mathrm{Planck}$ estimates are within $E(B-V)<0.3$\,mag and, hence, $A_\mathrm{V}<0.9$\,mag.

We first present a brief comparison among these maps themselves. This is illustrated in Figs.~\ref{fig:sfd98_gnilc.eps}--\ref{fig:sycplanck_dbb23.eps}, 
with additional pairs shown in Appendix\ref{addcmds}.
\footnote{\citetalias{schlaflyfinkbeiner2011} and \citetalias{chiang2023} are modifications of \citetalias{sfd98}. As shown in Figs.~\ref{fig:sfd98_sf11.eps} and \ref{fig:sfd98_csfd.eps} in Appendix\ref{addcmds}, the estimates from \citetalias{schlaflyfinkbeiner2011} are approximately equal to those of \citetalias{sfd98} scaled by a factor of 0.865, while the estimates from \citetalias{chiang2023} are nearly identical to those of \citetalias{sfd98}.}
For these comparisons, we use our compiled lists of galaxies and quasars, SN~Ia, Galactic globular clusters, and a subset of open clusters located behind the Galactic dust layer. This comparison shows that
\begin{itemize}
    \item All the figures demonstrate a clear segregation of the globular clusters located within the Galactic dust layer ($|b|<13\degr$, filled red circles) from all other objects --- galaxies and quasars, SN~Ia, open clusters, and globular clusters located behind the dust (SYC$_\mathrm{SFD98}$ and SYC$_\mathrm{Planck}$ cover only high and middle Galactic latitudes and, hence, provide no estimates for such a segregation). A comparison of Fig.~\ref{fig:sfd98_gnilc.eps} with Fig.~\ref{fig:sfd98_dbb23.eps}, or Fig.~\ref{fig:sf11_gnilc.eps} with Fig.~\ref{fig:sf11_dbb23.eps}, shows that \citetalias{gnilc} and \citetalias{dbb23} exhibit opposite segregation trends, resulting in the strongest contrast in Fig.~\ref{fig:gnilc_dbb23.eps}, where these maps are compared directly.
    A comparison of Figs.~\ref{fig:sfd98_gnilc.eps} and \ref{fig:sf11_gnilc.eps} shows that \citetalias{gnilc} agrees with \citetalias{schlaflyfinkbeiner2011} for extinction estimates toward globular clusters within the dust, while with \citetalias{sfd98} for estimates toward galaxies and SN~Ia. This indicates that no single 2D map can be considered superior in all regimes.
    This segregation may be explained by the spatial distribution and environment of the objects. Most globular clusters within the dust layer are located in the Galactic bulge near the Galactic center, while galaxies and SN~Ia are typically observed at middle and high Galactic latitudes, through the Galactic halo. Dust in these different regions may exhibit variations in properties such as temperature and extinction law \citep{gontcharov2013halo,gontcharov2013disk,planck2014,gontcharov2016,legnardi2023}. These variations may be perceived differently by the telescopes that provided the data for the respective maps --- {\it Gaia} for \citetalias{dbb23}, {\it COBE}+{\it IRAS} for \citetalias{sfd98}, and {\it Planck} for \citetalias{gnilc}.
    \item \citetalias{dbb23} additionally shows a segregation between globular clusters located behind the dust, on the one side, versus the galaxies, SN~Ia, and open clusters, on the other side. This segregation is difficult to explain by dust property variations alone, since these objects generally occupy similar regions of the sky ---especially in the Galactic halo --- and are often close neighbors. For instance, the globular cluster NGC\,6205 and SN~Ia PTF~11kqm are only 36 arcmin apart, yet \citetalias{dbb23} gives significantly different extinction estimates: $A_\mathrm{V} = 0.5$ and $0.3$~mag, respectively. These are much higher than the more consistent estimates from \citetalias{sfd98} ($A_\mathrm{V} = 0.06$~mag) and \citetalias{gnilc} ($A_\mathrm{V} = 0.05$~mag).
    Furthermore, Figs.~\ref{fig:gnilc_dbb23.eps}--\ref{fig:sf11_dbb23.eps} show that the \citetalias{dbb23} estimates exhibit a larger scatter compared to the other maps. Notably, extinction values in large areas around the Magellanic Clouds are strongly overestimated in the \citetalias{dbb23} map, as seen in the inset of Fig.~\ref{fig:sfd98_dbb23.eps}, even though we have excluded all objects physically associated with the Magellanic Clouds. These features suggest that the \citetalias{dbb23} estimates may be less reliable than those from other maps. However, a detailed analysis of this issue is beyond the scope of the present paper.
    \item SYC$_\mathrm{SFD98}$ and SYC$_\mathrm{Planck}$ are not much different from \citetalias{sfd98} and \citetalias{gnilc}, respectively. Some outliers among SYC$_\mathrm{SFD98}$ estimates are evident in Figs~\ref{fig:sfd98_syc.eps} and \ref{fig:sf11_sycsfd98.eps}. We have established that they are objects closest to the Galactic centre among all the objects in the high- and middle-latitude area covered by SYC$_\mathrm{SFD98}$. Therefore, these outliers are probably due to some spatial variations of extinction law.
\end{itemize}

\begin{figure} 
   \centering
   \includegraphics[width=14.0cm, angle=0]{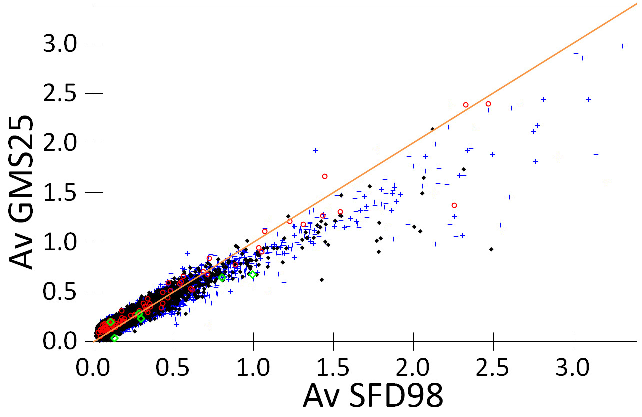}
\caption{$A_\mathrm{V}$ from \citetalias{sfd98} vs.\ our 2D map for objects located behind the Galactic dust layer ($|b|>13\degr$): 
galaxies and quasars (blue crosses), SN~Ia (black diamonds), open clusters (open green diamonds), and Galactic globular clusters (open red circles).
The orange line indicates the one-to-one relation.}
   \label{fig:sfd98_gms25.eps}
   \end{figure}

\begin{figure} 
   \centering
   \includegraphics[width=14.0cm, angle=0]{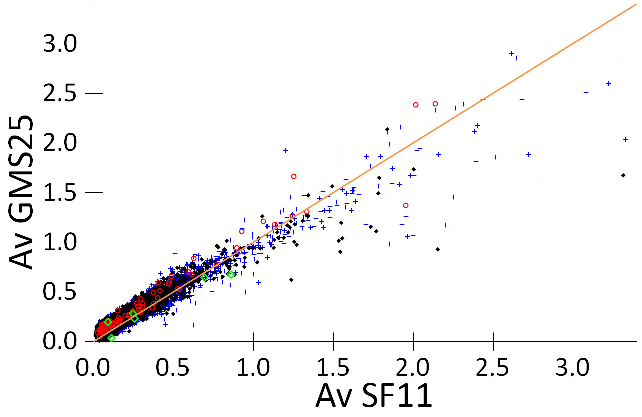}
   \caption{The same as Fig.~\ref{fig:sfd98_gms25.eps} but for \citetalias{schlaflyfinkbeiner2011} vs our 2D map.}
   \label{fig:sf11_gms25.eps}
   \end{figure}

\begin{figure} 
   \centering
   \includegraphics[width=14.0cm, angle=0]{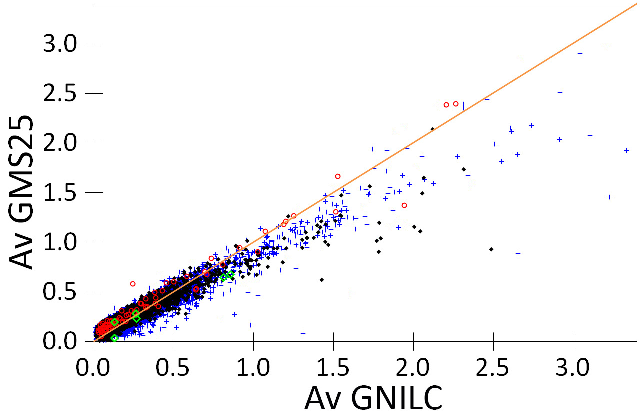}
   \caption{The same as Fig.~\ref{fig:sfd98_gms25.eps} but for \citetalias{gnilc} vs our 2D map.}
   \label{fig:gnilc_gms25.eps}
   \end{figure}

\begin{figure} 
   \centering
   \includegraphics[width=14.0cm, angle=0]{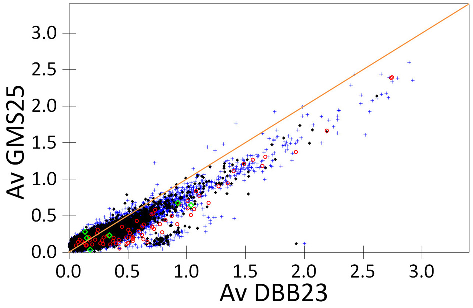}
   \caption{The same as Fig.~\ref{fig:sfd98_gms25.eps} but for \citetalias{dbb23} vs our 2D map.}
   \label{fig:dbb23_gms25.eps}
   \end{figure}

\begin{figure} 
   \centering
   \includegraphics[width=14.0cm, angle=0]{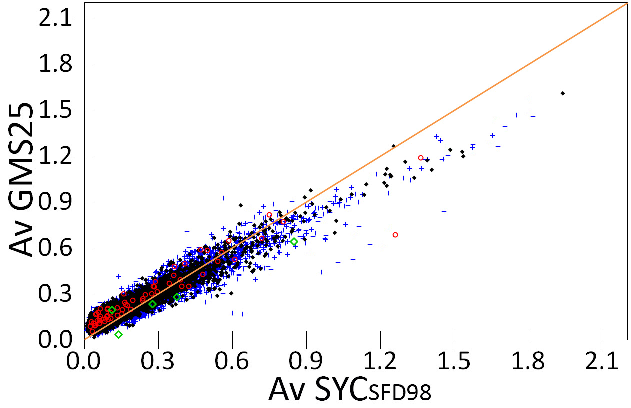}
   \caption{The same as Fig.~\ref{fig:sfd98_gms25.eps} but for SYC$_\mathrm{SFD98}$ vs our 2D map.}
   \label{fig:sycsfd98_gms25.eps}
   \end{figure}

\begin{figure} 
   \centering
   \includegraphics[width=14.0cm, angle=0]{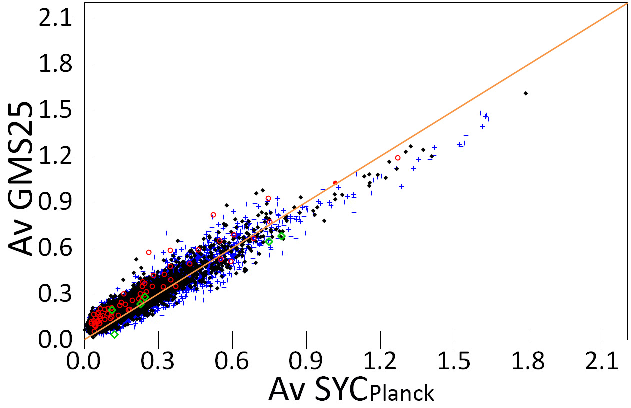}
   \caption{The same as Fig.~\ref{fig:sfd98_gms25.eps} but for SYC$_\mathrm{Planck}$ vs our 2D map.}
   \label{fig:sycplanck_gms25.eps}
   \end{figure}

\begin{figure} 
   \centering
   \includegraphics[width=14.0cm, angle=0]{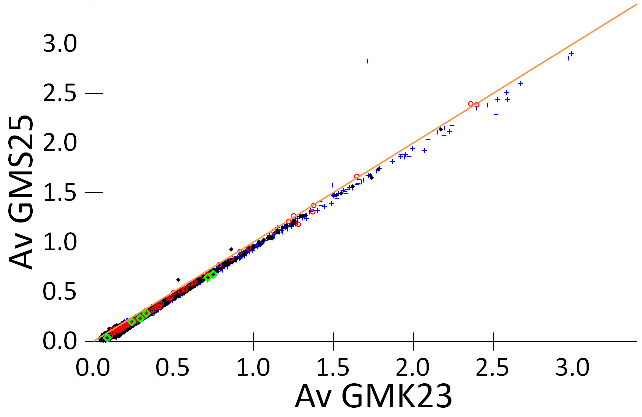}
   \caption{The same as Fig.~\ref{fig:sfd98_gms25.eps} but for our previous \citetalias{gontcharov2023} and current 2D maps.}
   \label{fig:gmk23_gms25.eps}
   \end{figure}

\begin{figure} 
   \centering
   \includegraphics[width=14.0cm, angle=0]{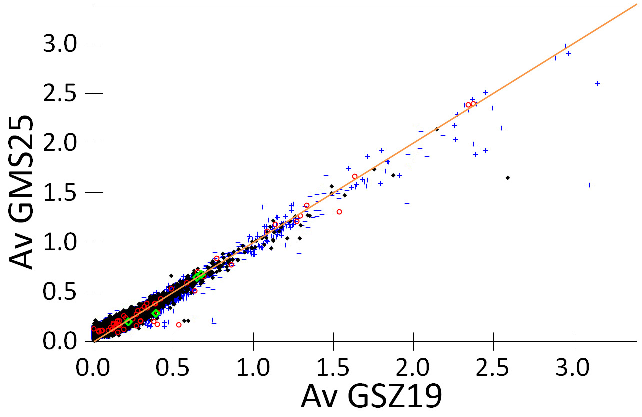}
   \caption{The same as Fig.~\ref{fig:sfd98_gms25.eps} but for \citetalias{green2019} vs our 2D map.}
   \label{fig:gsz19_gms25.eps}
   \end{figure}

Our 2D map predictions for the samples restricted to $|b|>13\degr$ is compared with those from \citetalias{sfd98}, \citetalias{schlaflyfinkbeiner2011}, \citetalias{gnilc}, \citetalias{dbb23}, SYC$_\mathrm{SFD98}$, and SYC$_\mathrm{Planck}$
in Figs~\ref{fig:sfd98_gms25.eps}--\ref{fig:sycplanck_gms25.eps}, respectively.
In addition, we compare our 2D map with the previous version of our 2D map from \citetalias{gontcharov2023} in Fig.~\ref{fig:gmk23_gms25.eps}, and with the farthest-distance-bin estimates from \citetalias{green2019} (interpreted here as a 2D map) in Fig.~\ref{fig:gsz19_gms25.eps}.

The map-to-map trends, calculated via least-squares fitting for 73 globular clusters at $|b|>13\degr$, are given below (hereafter we refer to our map as GMS25):
\begin{align*}
\label{terms}
\mathrm{GMS25} &= 0.995 \cdot \mathrm{GMK23} - 0.034 \\
\mathrm{GMS25} &= 0.957 \cdot \mathrm{GSZ19} + 0.008 \\
\mathrm{GMS25} &= 0.797 \cdot \mathrm{DBB23} - 0.117 \\
\mathrm{GMS25} &= 0.850 \cdot \mathrm{SFD98} + 0.073 \\
\mathrm{GMS25} &= 0.854 \cdot \mathrm{CSFD} + 0.072 \\
\mathrm{GMS25} &= 0.751 \cdot \mathrm{SYC_\mathrm{SFD98}} + 0.105 \\
\mathrm{GMS25} &= 0.983 \cdot \mathrm{SF11} + 0.073 \\
\mathrm{GMS25} &= 0.909 \cdot \mathrm{GNILC} + 0.078 \\
\mathrm{GMS25} &= 0.948 \cdot \mathrm{SYC_\mathrm{Planck}} + 0.091 \\
\end{align*}
Similar trends are found for galaxies and SN~Ia at $|b|>13\degr$, when limited to $A_\mathrm{V} < 2.7$~mag. This limit is justified, as higher extinctions are associated with increased uncertainties due to differences in map resolution, nonlinearities in reddening-to-extinction conversion, steep extinction gradients, and the obscuration of faint stars in dense dust clouds.

These trends indicate that our new 2D map (GMS25) yields values lower by $\Delta A_\mathrm{V} \approx 0.034$~mag compared to its previous version \citetalias{gontcharov2023}, is in good agreement with \citetalias{green2019}, and significantly diverges from the \citetalias{dbb23} map --- raising concerns about the reliability of the latter. The trends also reflect the similarity between \citetalias{sfd98} and \citetalias{chiang2023}, as seen in Fig.~\ref{fig:sfd98_csfd.eps}, and show comparable constant terms but different scaling coefficients when compared to \citetalias{sfd98}, \citetalias{schlaflyfinkbeiner2011}, and \citetalias{gnilc}.
SYC$_\mathrm{SFD98}$ and SYC$_\mathrm{Planck}$ show rather large constant terms w.r.t. our map, while their scaling coefficients differ significantly generally making them similar to \citetalias{sfd98} and \citetalias{gnilc}, respectively.

Our 2D map estimates are systematically higher than those from \citetalias{sfd98} and \citetalias{gnilc} in regions of low extinction, and lower in regions of high extinction. A notable zero-point offset is observed between our estimates and those from \citetalias{sfd98}, \citetalias{gnilc}, \citetalias{schlaflyfinkbeiner2011}, and \citetalias{chiang2023}, with a nearly constant difference of $\Delta A_\mathrm{V} \approx 0.07$--$0.08$~mag. 

It is well known that the maps by \citetalias{sfd98} and \citetalias{gnilc} tend to underestimate low and overestimate high reddening/extinction values \citep[][and references therein]{wolf2014,sun2022}. This systematic trend is confirmed by our 2D map, as demonstrated in Figs.~\ref{fig:sfd98_gms25.eps} and \ref{fig:gnilc_gms25.eps}. 
This trend is seen also for \citetalias{chiang2023}, SYC$_\mathrm{SFD98}$ and, to a lesser extent, for SYC$_\mathrm{Planck}$.
The modification of the \citetalias{sfd98} map proposed by \citetalias{schlaflyfinkbeiner2011} was specifically designed to mitigate this bias. The comparison of Figs~\ref{fig:sfd98_gms25.eps} and \ref{fig:sf11_gms25.eps} shows that this correction is effective. However, it introduces a significant zero-point offset, manifested as a large constant difference between the \citetalias{schlaflyfinkbeiner2011} estimates and those from our map.

The 2D maps of \citetalias{sfd98}, its modification by \citetalias{schlaflyfinkbeiner2011}, as well as \citetalias{gnilc} are based on observations of the infrared emission from interstellar dust. For their emission-to-reddening calibration, \citetalias{sfd98} use 389 elliptical galaxies, \citetalias{schlaflyfinkbeiner2011} use 261,496 stars, and \citetalias{gnilc} adopt the calibration from \citet{planck2014}, which employs 53,399 quasars. In contrast, our 2D map is constructed using approximately 35 million dwarf stars located within 450--3000~pc from the Sun (depending on Galactic latitude), i.e. over a hundred times more objects than used by \citetalias{schlaflyfinkbeiner2011}. 
This large sample size represents a key advantage of our approach over previous 2D maps. Another advantage is the significantly improved representation of stellar SEDs enabled by multi-band photometry from {\it Gaia}, PS1, SMSS, 2MASS, and {\it WISE}, as used by \citetalias{anders2022}. Moreover, we have made substantial efforts to suppress systematic errors in the input data, as demonstrated in Sect.~\ref{sect:systematics}. The very detection of systematics in $A_\mathrm{V}$ at the level of a few hundredths of a magnitude indicates that the systematic accuracy of our map is likely at least this good.

We therefore conclude that our 2D map seems to be systematically more accurate than the emission-based 2D maps under consideration. Consequently, some of the systematic differences between our map and the others should be interpreted as manifestations of systematics in those earlier maps.

Nevertheless, the accuracy of our 2D map requires further verification. The lists of galaxies, quasars, SN~Ia, and star clusters compiled in this study can serve as useful benchmarks for such validation in the future.

A noticeable difference between the maps appears in the estimates of the TGE across the entire dust half-layer below or above the Sun, averaged over the SGP and NGP. 
The corresponding values of $A_\mathrm{V}$ are 
0.093,    0.086,             0.043,              0.041,                0.037,                               0.039, and     0.036\,mag from 
our map, \citetalias{dbb23}, \citetalias{sfd98}, SYC$_\mathrm{SFD98}$, \citetalias{schlaflyfinkbeiner2011}, \citetalias{gnilc}, and SYC$_\mathrm{Planck}$ respectively. The difference between our TGE estimate and those from \citetalias{sfd98}, \citetalias{schlaflyfinkbeiner2011}, and \citetalias{gnilc} vastly contributes the constant term (about 0.07\,mag) of their systematic difference all over the sky. Furthermore, taking into account rather high uncertainty of the TGE estimates\footnote{For example, \citetalias{sfd98} state the standard deviation $\sigma E(B-V)=0.028$\,mag, that is $\sigma(A_\mathrm{V})\approx0.09$~mag, of their estimates, albeit may be better after averaging for TGE.}, all of them, except ours and \citetalias{dbb23}, should be considered as insignificantly different from zero.

Nevertheless, some relative estimates are meaningful. For example, all these 2D maps indicate higher extinction toward the SGP than toward the NGP. For example, our map gives $A_\mathrm{V}=0.097$~mag at the SGP versus 0.089~mag at the NGP, yielding a difference $\Delta A_\mathrm{V}=0.008$~mag. This asymmetry reflects the Sun's position above the Galactic mid-plane and main concentration of the Galactic dust layer. Our estimate of this difference is in agreement with the values 
$\Delta A_\mathrm{V}=0.013$, 0.020,               0.011,                               0.009,              0.004, and 0.012~mag from 
\citetalias{sfd98}, SYC$_\mathrm{SFD98}$, \citetalias{schlaflyfinkbeiner2011}, \citetalias{gnilc}, SYC$_\mathrm{Planck}$, and \citetalias{dbb23}, respectively.
Given that the Sun lies approximately 20~pc above the Galactic mid-plane \citep[][and references therein]{gontcharov2012ob,gontcharov2021a}, this suggests that non-zero extinction at the level of $A_\mathrm{V}\approx0.01$\,mag exists even in the immediate solar neighborhood within 40~pc from the Sun. Therefore, even this region cannot be considered dust-free when extinction at the 0.01~mag level is of interest. Currently, this can be considered as the desired level. Nevertheless, this makes it possible to understand why the \citet{sun2022} results are similar to those from \citetalias{sfd98} and {\it Planck}. \citet{sun2022} use a control sample of unreddened or negligibly reddened stars. This sample is created by several criteria including $|Z|>200$\,pc. This means that the authors ignore a significant fraction of the TGE occurring within $|Z|<200$\,pc. Therefore, \citet{sun2022} reproduce the insignificant very low TGE estimates from \citetalias{sfd98} or \citetalias{gnilc} and, consequently, underestimate low extinctions.
However, \citet{gontcharov2021a} analyze colours of a complete sample of 101\,810 red clump giants from {\it Gaia} DR2 and show that they can be considered as unreddened ones at a level of $E(B-V)<0.01$ or $A_\mathrm{V}<0.03$\,mag only within $|Z|<100$\,pc from the Sun. Else, considering slightly reddened stars as unreddened ones, one introduces a bias to any further estimate of the reddening for distant stars.

\section{Conclusions}
\label{sect:conclusions}

Despite significant progress in mapping Galactic extinction, further refinement and validation of existing maps remain essential. The considerable potential of {\it Gaia} data has not yet been fully realized. For instance, the distances, extinctions, and stellar parameters derived by \citetalias{anders2022}, based on {\it Gaia}~DR3 parallaxes and multi-band photometry for several hundred million stars, have been rarely utilized to determine the foreground extinction to specific celestial objects or to construct extinction maps. One key reason is the frequent occurrence of unphysical extinction trends, such as decreasing extinction with increasing distance along a LOS, which indicate the presence of significant systematic errors in the \citetalias{anders2022} estimates. As a result, extinction maps derived from these data, such as those by \citetalias{dbb23} and \citetalias{gontcharov2023}, often inherit these systematic issues and therefore have limited reliability and applicability.

In this study, we have identified and addressed key sources of systematic error in the extinction and distance estimates from \citetalias{anders2022}, successfully suppressing these systematics to within a few hundredths of a magnitude in $A_\mathrm{V}$. 
Based on the cleaned data, we have constructed new 2D and 3D extinction maps. Our analysis shows that correcting for three specific systematic effects is essential for improving the accuracy of $A_\mathrm{V}$ estimates:  
(i) a systematic deviation of isochrones from the data in the $M_G$--$A_\mathrm{V}$ plane, likely due to deficiencies in the modeling of low-mass stars;  
(ii) a systematic trend in $|Z|$--$A_\mathrm{V}$ space, likely caused by improper metallicity assumptions for the best-fitting isochrone, coupled with the well-known degeneracy between metallicity and extinction; and  
(iii) a bias in $A_\mathrm{V}$ due to the exclusion of nearby faint stars with low parameter fidelity.  
In constructing our new maps, we generally followed the methodology of \citetalias{gontcharov2023}, with several refinements. We used the $A_\mathrm{V}$ and distance estimates from \citetalias{anders2022} for nearly 100 million dwarf stars, corrected for the identified systematics. Special attention was given to the local solar neighborhood within 200~pc and to high Galactic latitudes.  
As with any reddening or extinction map, our maps inevitably smooth over small-scale fluctuations in the dust distribution, and are thus more reliable for estimating extinction to extended rather than point sources. Accordingly, we have compiled extensive catalogs of extended objects with angular sizes mainly in the range 2--40 arcmin for validation and testing, including 19,809 galaxies and quasars, 170 Galactic globular clusters, 458 open clusters, as well as a list of 8,293 Type Ia supernovae. Additionally, we have analyzed two sets of 318 and 537 molecular clouds from \citetalias{zucker2020} and \citetalias{chen2020}, respectively.

Our 2D map of the TGE $A_\mathrm{V}$ across the entire dust half-layer from the Sun to extragalactic space for Galactic latitudes $|b|>13\degr$ achieves a stated accuracy of $\sigma(A_\mathrm{V}) = 0.07$ mag and an angular resolution of 6.1 arcmin --- matching that of \citetalias{sfd98}, \citetalias{schlaflyfinkbeiner2011}, and \citetalias{chiang2023}. We have validated our 2D map by comparing its extinction estimates with those from other 2D maps, as well as with literature values for 45 Galactic globular clusters located behind the Galactic dust layer and within 25 kpc of the Sun. These comparisons reveal systematic inconsistencies in the predictions of all 2D maps, on the order of several hundredths of a magnitude in $A_\mathrm{V}$. Such discrepancies can likely be attributed to large-scale spatial variations in the extinction law and to systematic errors in the emission-to-reddening calibrations adopted by some maps. Given that our 2D map is based on an unprecedented number of stars and benefits from superior stellar SED representation using multi-band photometry, we argue that it is among the most reliable currently available and is well suited for further investigation of systematic uncertainties in extinction mapping.

Our 3D map of $A_\mathrm{V}$ within 2~kpc of the Sun features a transverse resolution of 3.56~pc and a radial resolution of 50~pc. It provides estimates of $A_\mathrm{V}$ from the Sun to extended objects embedded within the Galactic dust layer with an accuracy of $\sigma(A_\mathrm{V}) = 0.1$~mag. We have validated our 3D map by comparing its predictions with those from the 3D map of \citetalias{green2019}, as well as with literature estimates for globular clusters within the Galactic dust layer, open clusters, and molecular clouds from the selected samples. The results show good agreement, demonstrating the utility of our map for determining the distance, foreground extinction, and backside extinction of extended objects located within the Galactic dust in future studies.

\normalem
\begin{acknowledgements}
We acknowledge financial support from the Russian Science Foundation (grant no. 20--72--10052).

We thank the referee for a constructive and very useful report.
We thank
Gregory Green for discussion of his extinction/reddening estimates,
Maxim Khovritchev for technical support,
and Eugene Vasiliev for his very useful comments on the globular cluster properties.

This work has made use of
Filtergraph \citep{filtergraph}, an online data visualization tool developed at Vanderbilt University through the Vanderbilt Initiative in 
Data-intensive Astrophysics (VIDA) and the Frist Center for Autism and Innovation (FCAI, \url{https://filtergraph.com});
the resources of the Centre de Donn\'ees astronomiques de Strasbourg, Strasbourg, France (\url{http://cds.u-strasbg.fr}), including the SIMBAD database \citep{simbad}, 
the VizieR catalogue access tool \citep{vizier} and the X-Match service;
data from the European Space Agency (ESA) mission {\it Gaia} (\url{https://www.cosmos.esa.int/gaia}), processed by the {\it Gaia} Data Processing and Analysis Consortium 
(DPAC, \url{https://www.cosmos.esa.int/web/gaia/dpac/consortium}), and {\it Gaia} archive website (\url{https://archives.esac.esa.int/gaia});
the HyperLeda database (\url{http://leda.univ-lyon1.fr});
the NASA/IPAC Extragalactic Database, which is funded by the National Aeronautics and Space Administration and operated by the California Institute of Technology.

\end{acknowledgements}

\bibliographystyle{raa}
\bibliography{bibtex}

\clearpage
\appendix

\section{Additional comparison of various maps}
\label{addcmds}

\begin{figure}[h] 
   \centering
   \includegraphics[width=13.0cm, angle=0]{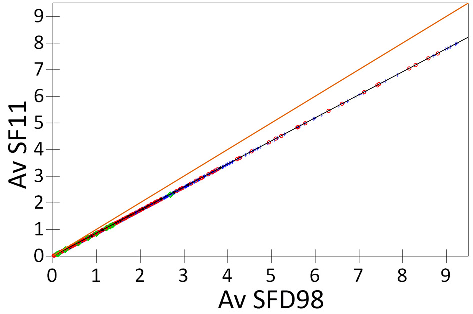}
\caption{$A_\mathrm{V}$ from \citetalias{sfd98} vs \citetalias{schlaflyfinkbeiner2011} for galaxies and quasars (blue crosses), SN~Ia (black diamonds), open clusters behind the Galactic dust layer (open green diamonds), Galactic globular clusters behind the layer (open red circles), and globular clusters within the layer (filled red circles). The orange line shows the one-to-one relation. The black line indicates the trend with a coefficient of 0.865.}
   \label{fig:sfd98_sf11.eps}
   \end{figure}

\begin{figure}[h]
   \centering
   \includegraphics[width=13.0cm, angle=0]{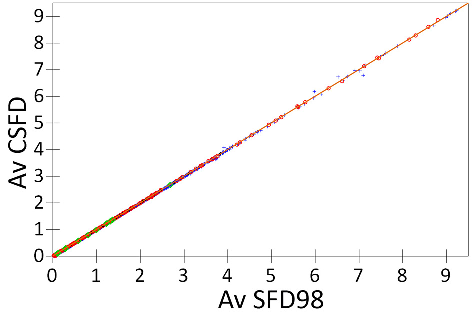}
   \caption{The same as Fig.~\ref{fig:sfd98_sf11.eps} but for \citetalias{sfd98} vs \citetalias{chiang2023}.}
   \label{fig:sfd98_csfd.eps}
   \end{figure}

\begin{figure}[h]
   \centering
   \includegraphics[width=13.0cm, angle=0]{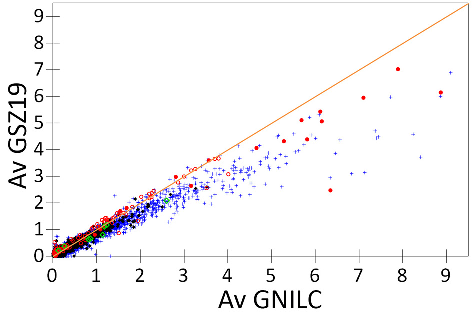}
   \caption{The same as Fig.~\ref{fig:sfd98_sf11.eps} but for \citetalias{gnilc} vs \citetalias{green2019}.}
   \label{fig:gnilc_gsz19.eps}
   \end{figure}

\begin{figure}[h]
   \centering
   \includegraphics[width=13.0cm, angle=0]{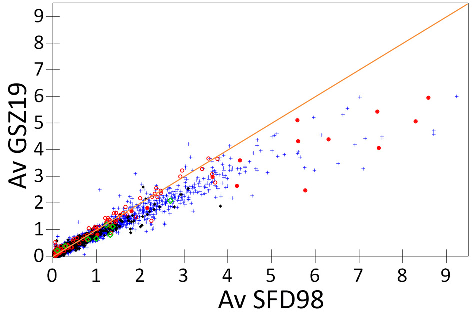}
   \caption{The same as Fig.~\ref{fig:sfd98_sf11.eps} but for \citetalias{sfd98} vs \citetalias{green2019}.}
   \label{fig:sfd98_gsz19.eps}
   \end{figure}

\begin{figure}[h]
   \centering
   \includegraphics[width=13.0cm, angle=0]{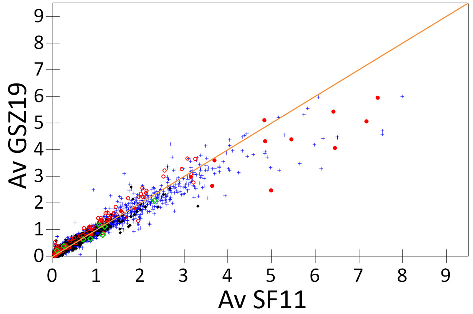}
   \caption{The same as Fig.~\ref{fig:sfd98_sf11.eps} but for \citetalias{schlaflyfinkbeiner2011} vs \citetalias{green2019}.}
   \label{fig:sf11_gsz19.eps}
   \end{figure}

\begin{figure}[h]
   \centering
   \includegraphics[width=13.0cm, angle=0]{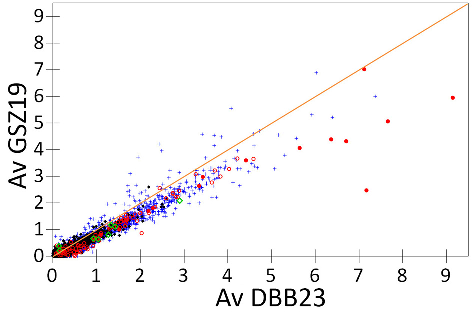}
   \caption{The same as Fig.~\ref{fig:sfd98_sf11.eps} but for \citetalias{dbb23} vs \citetalias{green2019}.}
   \label{fig:dbb23_gsz19.eps}
   \end{figure}

\begin{figure}[h]
   \centering
   \includegraphics[width=13.0cm, angle=0]{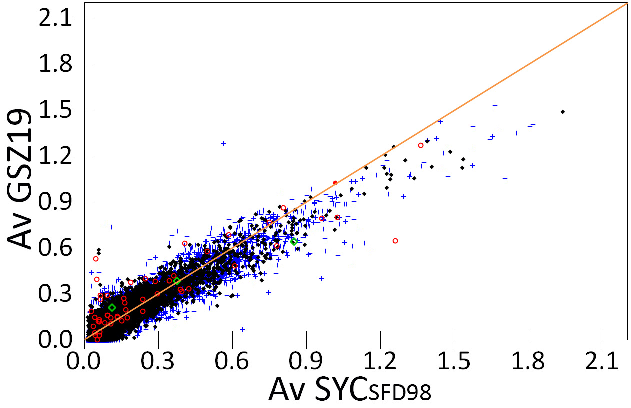}
   \caption{The same as Fig.~\ref{fig:sfd98_sf11.eps} but for SYC$_\mathrm{SFD98}$ vs \citetalias{green2019}.}
   \label{fig:sycsfd98_gsz19.eps}
   \end{figure}

 \begin{figure}[h]
   \centering
   \includegraphics[width=13.0cm, angle=0]{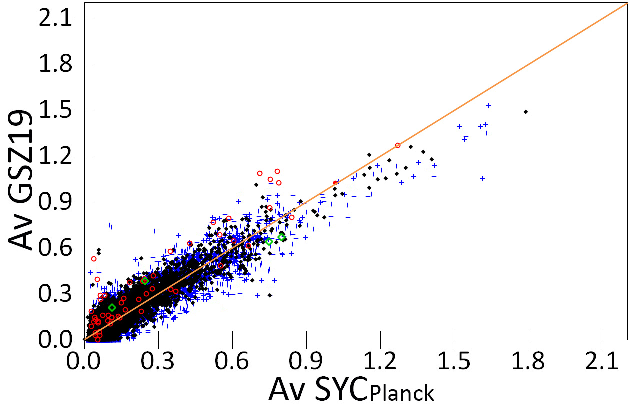}
   \caption{The same as Fig.~\ref{fig:sfd98_sf11.eps} but for SYC$_\mathrm{Planck}$ vs \citetalias{green2019}.}
   \label{fig:sycplanck_gsz19.eps}
   \end{figure}

\end{document}